\documentclass[useAMS]{gCOM2e}

\usepackage{amsmath}
\usepackage{amssymb}

\newcounter{step}[subsection]
\newcommand{\step}[1]{\textit{Step \stepcounter{step}\thestep~#1.}}
\renewcommand{\thestep}{\textit{\arabic{step}}}
\newcounter{substep}[step]
\renewcommand{\thesubstep}{\textit{\arabic{step}.\arabic{substep}}}
\newcommand{\substep}[1]{\textit{Step \stepcounter{substep}\thesubstep~#1.}}
\renewcommand{\vec}[1]{\mathbf{#1}}

\DeclareMathOperator{\rank}{rank}
\newcommand{\R}{\mathcal{R}}
\newcommand{\Euler}[1]{\mathcal{L}_{#1}}

\title{A symbolic algorithm for computing recursion operators of 
nonlinear PDEs}

\author{%
  D.E.\ Baldwin$^{\rm \dagger}$$^{\ast}$%
    \thanks{$^\ast$Corresponding author. %
      Email: recursionOperators@douglasbaldwin.com
      \vspace{6pt}} and 
  W.\ Hereman$^{\rm \ddagger}$\\\vspace{6pt}
  $^{\rm \dagger}${\em{Department of Applied Mathematics, 
    University of Colorado, UCB 526, Boulder, CO 80309-0526, USA}}; 
  $^{\rm \ddagger}${\em{Department of Mathematical and Computer Sciences,
    Colorado School of Mines, Golden, CO 80401-1887, USA}}\\\vspace{6pt}
  \received{Received 31 March 2009}}

\begin{document}

\maketitle
%
%
\begin{abstract}

A recursion operator is an integro-differential operator which maps a 
generalized symmetry of a nonlinear PDE to a new symmetry.  
Therefore, the existence of a recursion operator guarantees that the PDE has 
infinitely many higher-order symmetries, which is a key feature of complete 
integrability. 
Completely integrable nonlinear PDEs have a bi-Hamiltonian structure and 
a Lax pair; they can also be solved with the inverse scattering transform 
and admit soliton solutions of any order. 

A straightforward method for the symbolic computation of polynomial 
recursion operators of nonlinear PDEs in $(1+1)$ dimensions is presented. 
Based on conserved densities and generalized symmetries, 
a candidate recursion operator is built from a linear combination 
of scaling invariant terms with undetermined coefficients.
The candidate recursion operator is substituted into its defining equation 
and the resulting linear system for the undetermined coefficients is solved.

The method is algorithmic and is implemented in {\it Mathematica}. 
The resulting symbolic package \texttt{PDE\-Re\-cur\-sion\-Op\-er\-a\-tor.m} 
can be used to test the complete integrability of polynomial PDEs that can 
be written as nonlinear evolution equations. 
With \texttt{PDE\-Re\-cur\-sion\-Op\-er\-a\-tor.m}, 
recursion operators were obtained for several well-known nonlinear PDEs 
from mathematical physics and soliton theory. 


\begin{classcode}
Primary: 37K10, 35Q51, 68W30;
Secondary: 37K05, 35Q53, 35Q58, 
\end{classcode}
\end{abstract}

%
%
\section{Introduction}
\label{sec:history}

Completely integrable nonlinear partial differential equations (PDEs) have a 
rich mathematical structure and many hidden properties. 
For example, these PDEs have infinitely many conservation laws and 
generalized symmetries of increasing order.
They have the Painlev\'e property \cite{Ablowitz80a}, bi-Hamiltonian 
(sometimes tri-Hamiltonian) structures \cite{AblowitzSegur81}, 
Lax pairs \cite{Ablowitz91}, 
B\"ack\-lund and Dar\-boux transformations \cite{Lakshmanan83,Olver93}, etc.
Completely integrable PDEs can be solved with the Inverse Scattering Transform
(IST) \cite{Ablowitz91,AblowitzSegur81,CalogeroDegasperis82}. 
Application of the IST or Hirota's direct method \cite{Hirota80,Hirota04} 
allows one to construct explicit soliton solutions of any order. 
While there are numerous definitions of complete integrability, 
Fokas~\cite{Fokas87} defines an equation as completely integrable if and only 
if it possesses infinitely many generalized symmetries.  
A \emph{recursion operator} (also called a formal symmetry or a 
master symmetry) is a linear integro-differential operator which links 
such symmetries.
The recursion operator is thus a key tool for proving the existence of an 
infinite hierarchy of generalized symmetries~\cite{Olver93} and for 
computing them sequentially.   


The recursion operator story \cite{Olver93,Wang98} starts with the 
Korteweg-de Vries (KdV) equation,
\begin{equation}
  \label{kdv}
  u_t + 6 u u_x + u_{3x} = 0,
\end{equation}
which is undeniably the most famous completely integrable PDE.
The first few generalized symmetries (of infinitely many) for the KdV 
equation are 
\begin{equation}
  \label{kdvSym}
  \begin{gathered}
    G^{(1)} = u_x, \qquad G^{(2)} = 6uu_x + u_{3x}, \\
    G^{(3)} = 30u^2 u_x + 20 u_x u_{2x} + 10 u u_{3x} + u_{5x}.
  \end{gathered}
\end{equation}
Note that generalized symmetries depend on the dependent variables of the 
system as well as the $x$-derivatives of the dependent variables 
(in contrast to so-called Lie-point or {\em geometric} symmetries which 
only depend on the independent and dependent variables of the system). 

The KdV equation is a member of a hierarchy of integrable PDEs, which are 
higher-order symmetries of the KdV itself. 
For example, the Lax equation \cite{Lax68}, 
which is the fifth-order member in the hierarchy, is $u_t + G^{(3)} = 0.$ 

Based on the recursion formula \cite{PraughtSmirnov05} due to Lenard, 
in 1977 Olver \cite{Olver77} derived an explicit recursion operator for 
the KdV equation, namely
\begin{equation}
  \label{kdvRecursion}
  \R = D_x^2 + 4 u I + 2 u_x D_x^{-1}.
\end{equation}
In (\ref{kdvRecursion}), $D_x$ denotes the total derivative with respect to 
$x,$ $D_x^{-1}$ is its left inverse, and $I$ is the identity operator.
Total derivatives act on differential functions \cite{Olver93}, 
i.e.\ differentiable functions of independent variables, dependent variables, 
and their derivatives up to an arbitrary but fixed order.

The recursion operator (\ref{kdvRecursion}) allows one to generate an 
infinite sequence of local generalized symmetries of the KdV equation.
Indeed, starting from ``seed" or ``root" $G^{(1)},$ repeated application 
of the recursion operator (\ref{kdvRecursion}), 
\begin{equation}
\label{sequentialsymmetries}
G^{(j+1)} = \R G^{(j)}, \qquad j = 1,2,\ldots ,
\end{equation}
produces the symmetries in (\ref{kdvSym}) and infinitely many more.

Analysis of the form of recursion operators like (\ref{kdvRecursion}) 
reveals that they can be split into a (local) differential part, $\R_0,$ 
and a (non-local) integral part $\R_1.$  
The differential operator $\R_0$ involves $D_x, D_x^{2},$ etc., 
acting on monomials in the dependent variables. 
Barring strange cases \cite{Karasu04}, the integral operator $\R_1$ only 
involves $D_x^{-1}$ and can be written as the outer product of 
generalized symmetries and cosymmetries 
(or conserved covariants)~\cite{Bilge93,Wang98}.  
Furthermore, if $\R$ is a recursion operator, then the 
Lie derivative~\cite{Olver93,Wang98,Wang02} of $\R$ with respect to 
the evolution equation is zero.  
The latter provides an explicit defining equation for the recursion operator. 

For more information on the history of completely integrable systems 
and recursion operators, 
see~\cite{Ablowitz91,Bilge93,Dodd82,Dorfman93,Fokas87,Konopelchenko87,%
Lamb80,Olver93,Fuchssteiner82,Fuchssteiner87,Fuchssteiner97,%
MikhailovShabatSokolov91,Newell85,Sanders98,Sanders01,Sanders01bis,Wang98}.  
Based on studies of formal symmetries and recursion operators, researchers 
have compiled lists of integrable PDEs 
\cite{MikhailovShabatYamilov87,MikhailovShabatSokolov91,Wang98,Wang02}. 

While the computation of the Lie derivative of $\R$ is fairly straightforward, 
it is computationally intensive and prone to error when done by hand. 
For example, the computation of the recursion operators of the 
Kaup-Kupershmidt equation or the Hirota-Satsuma system 
(see Section~\ref{sec:examples}) 
may take weeks to complete by hand and might fill dozens of pages. 
Using a computer algebra system to carry out the lengthy computations 
is recommended, yet, it is not without challenge either; 
systems like {\it Mathematica} and {\it Maple} are designed to 
primarily work with commutative algebraic structures and the computation 
and application of recursion operators requires non-commutative operations.

There is a variety of methods \cite{Meshkov00} to construct recursion  
operators (or master symmetries).
As shown in \cite{Fokas87,FokasFuchssteiner80,FuchssteinerFokas81,Olver93}, 
one first finds a bi-Hamiltonian structure (with Hamiltonians $\Theta_1$ 
and $\Theta_2$) for the given evolution equation and then constructs 
the recursion operator as $\mathcal{R} = \Theta_1 \Theta_2^{-1}$, 
provided $\Theta_2$ is invertible; 
for the KdV equation, $\Theta_1 = D_x^3 + 4 u D_x + 2 u_x I$ 
and $\Theta_2 = D_x$ form a Hamiltonian pair \cite{magri78} and 
$\mathcal{R} = \Theta_1 \Theta_2^{-1}$ yields (\ref{kdvRecursion}).
The Hamiltonians are cosymplectic operators, their inverses are symplectic 
operators~\cite{Wang98}. 
A complicated example of a recursion operator (obtained by composing 
cosymplectic a symplectic operators of a vector derivative Schr\"odinger 
equation) can be found in \cite{Willoxetal95}.

A recent approach \cite{MikhailovNovikov02} uses the symbolic method of 
Gelfand and Dickey \cite{Gelfand75}, and applies to non-local and 
non-evolutionary equations such as the Benjamin-Ono and Camassa-Holm 
equations. 

%

At the cost of generality, we advocate a direct approach which applies to 
polynomial evolution equations.
In the spirit of work by Bilge \cite{Bilge93}, we use the scaling invariance 
of the given PDE to build a polynomial candidate recursion operator as a 
linear combination of scaling homogeneous terms with constant undetermined 
coefficients.  
The defining equation for the recursion operator is then used to compute the 
undetermined coefficients. 

The goals of our paper are threefold. 
We present
(i) an algorithmic method in a language that appeals to specialists and 
non-specialists alike,  
(ii) a symbolic package in {\it Mathematica} to carry out the tedious 
computations, 
(iii) a set of carefully selected examples to illustrate the algorithm and 
the code.

The theory on which our algorithm is based has been covered extensively 
in the literature \cite{Bilge93,Olver93,Sanders01,Sanders01bis,Wang98,Wang02}.
Our paper focuses on {\it how} things work rather than on {\it why} they work.
%
%
%
%

The package 
\texttt{PDE\-Re\-cur\-sion\-Op\-er\-a\-tor.m}~\cite{recursioncode05}
is part of our symbolic software collection for the integrability testing of 
nonlinear PDEs, including algorithms and {\em Mathematica} codes 
for the Painlev\'e test \cite{DBthesis,DBandWHpaisoftware,baldwinhereman2006} 
and the computation of conservation laws 
\cite{PAthesis,PAandWHsoftware,HEthesis,HEandWHsoftware,UGandWHpd98,%
heremanijqc2006,WHetalbook08,WHetalcpc98,Heremanetal05}, 
generalized symmetries 
\cite{Goktas98,GoktasHereman99} 
and recursion operators \cite{Goktas98,BaldwinCRM2005}.
As a matter of fact, our package 
\texttt{PDE\-Re\-cur\-sion\-Op\-er\-a\-tor.m} builds on the code 
\texttt{In\-var\-i\-ants\-Sym\-me\-tries.m} \cite{Goktas97code} for the 
computation of conserved densities and generalized symmetries for 
nonlinear PDEs.
%
%
The code \texttt{PDE\-Re\-cur\-sion\-Op\-er\-a\-tor.m} automatically computes 
{\em polynomial} recursion operators for polynomial systems of nonlinear 
PDEs in $(1+1)$ dimensions, i.e.\ PDEs in one space variable $x$ and time $t.$
At present, the coefficients in the PDEs cannot {\it explicitly} depend on 
$x$ and $t.$
Our code can find recursion operators with coefficients that {\em explicitly} 
depend on powers of $x$ and $t$ as long as the maximal degree of these 
variables is specified.
For example, if the maximal degree is set to $1$, then the coefficients will 
be at most linear in both $x$ and $t.$
An example of a recursion operator that explicitly depends on $x$ and 
$t$ is given in Section~\ref{subsec:burgers}.
For extra versatility, the code can be used to test polynomial and rational 
recursion operators found in the literature, computed by hand, or with 
other software packages.  

Drawing on the analogies with the PDE case, we also developed methods, 
algorithms, and software to compute conservation laws 
\cite{UGetalpla97,WHetalbook08,WHetalcpc98,MHandWHprsa03} and 
generalized symmetries 
\cite{GoktasHereman99}
of nonlinear differential-difference equations (DDEs).
Although the algorithm is well-established \cite{WHetalcrm05}, 
a {\em Mathematica} package that automatically computes recursion operators 
of nonlinear DDEs is still under development.

The paper is organized as follows.
In Section~\ref{sec:algorithm} we briefly discuss our method for computing 
scaling invariance, conserved densities, and generalized symmetries 
(which are essential pieces for the computation of recursion operators).  
Our method for computing and testing recursion operators is discussed in 
Section~\ref{sec:algoRecursion}.
In Section~\ref{sec:examples}, we illustrate the subtleties of the method 
using the KdV equation, the Kaup-Kupershmidt equation, and the 
Hirota-Satsuma system of coupled KdV (cKdV) equations.
The details of computing and testing recursion operators are discussed in 
Section~\ref{sec:keyAlgorithms}.  
Section~\ref{sec:otherSoftware} compares our software package to other 
software packages for computing recursion operators.  
In Section~\ref{sec:additionalExamples} we give additional examples to
demonstrate the capabilities of our software.  
A discussion of the results and future generalizations are given in 
Section~\ref{sec:conclusions}.
The use of the software package 
\texttt{PDE\-Re\-cur\-sion\-Op\-er\-a\-tor.m}~\cite{recursioncode05} 
is shown in Appendix~\ref{sec:softwareUsage}.  

%
%
\section{Scaling Invariance, Conservation Laws, and Generalized Symmetries}
\label{sec:algorithm}
Consider a polynomial system of evolution equations in $(1+1)$ dimensions,
\begin{equation}
  \label{PDESystem}
  \vec{u}_t(x,t) = \vec{F}(\vec{u}(x,t), \vec{u}_x(x,t), \vec{u}_{2x}(x,t), 
    \dotsc, \vec{u}_{mx}(x,t)), 
\end{equation}
where $\vec{F}$ has $M$ components $F_1$, \dots, $F_M$, 
$\vec{u}(x,t)$ has $M$ components $u_1(x,t)$, \dots, $u_M(x,t)$ and 
$\vec{u}_{ix} = {\partial^i \vec{u}}/{\partial x^i}$.
Henceforth we write $\vec{F}(\vec{u})$ although $\vec{F}$ (typically) 
depends on $\vec{u}$ and its $x$-derivatives up to some fixed order $m.$  
In the examples, we denote the components of $\vec{u}(x,t)$ as 
$u$, $v$, \dots, $w$.
If present, any parameters in the PDEs are assumed to be nonzero and 
are denoted by Greek letters.   

Our algorithms are based on scaling (or dilation) invariance, 
a feature common to many nonlinear PDEs. 
If (\ref{PDESystem}) is scaling invariant, then quantities like 
conserved densities, fluxes, generalized symmetries, and recursion operators 
are also scaling invariant~\cite{Olver93}.  
Indeed, since their defining equation must be satisfied on solutions of 
the PDE, these quantities ``inherit'' the scaling symmetry of the 
original PDE. 
Thus, scaling invariance provides an elegant way to construct the form of 
candidate densities, generalized symmetries, and recursion operators.
It suffices to make linear combinations (with constant undetermined 
coefficients) of scaling-homogeneous terms.
Inserting the candidates into their defining equations then leads to a linear 
system for the undetermined coefficients.

%
%
\subsection{Scaling Invariance and the Computation of Dilation Symmetries}
\label{subsec:scalinginvariance}

Many completely integrable nonlinear PDEs are scaling invariant.
PDEs that are not scaling invariant can be made so by extending the set 
of dependent variables with parameters that scale appropriately, 
see \cite{Goktas97,GoktasHereman99} for details.  

For example, the KdV equation (\ref{kdv}) is invariant under the 
scaling symmetry
\begin{equation}
  \label{kdvscalingsymmetry}
  (t,x,u) \to (\lambda^{-3} t, \lambda^{-1} x, \lambda^2 u),
\end{equation}
where $\lambda$ is an arbitrary parameter.  
Indeed, upon scaling, a common factor $\lambda^5$ can be pulled out.
Assigning \emph{weights} (denoted by $W)$ to the variables based on the 
exponents in $\lambda$ and setting $W(D_x) = 1$ 
(or equivalently, $W(x) = W(D_x^{-1}) = -1$)
gives $W(u) = 2$ and $W(t) = -3$ (or $W(D_t) = 3$).

The \emph{rank} of a monomial is its total weight; 
in the KdV equation, all three terms are rank $5$.
We say that an equation is \emph{uniform in rank} if every term in the 
equation has the same rank.
Conversely, requiring uniformity in rank in (\ref{kdv}) yields
\begin{equation}
 W(u) + W(D_t) = 2 W(u) + W(D_x) = W(u) + 3 W(D_x).
\end{equation}
Hence, after setting $W(D_x) = 1,$ 
one obtains $W(u) = 2 W(D_x) = 2$ and $W(D_t) = 3 W(D_x) = 3.$  
So, scaling symmetries can be computed with linear algebra. 

%
%
\subsection{Computation of Conservation Laws}
\label{subsec:conservationlaws}

The first two conservation laws for the KdV equation are 
\begin{gather}
  D_t( u ) + D_x( 3 u^2 + u_{2x} ) = 0, \\
  D_t \big( u^2 \big) + D_x \big( 4 u^3 - u_x^2 + 2 u u_{xx} \big) = 0, 
\end{gather}
were classically known and correspond to the conservation of mass and 
momentum (for water waves).
Whitham found the third conservation law,
\begin{equation}
  \label{whitham}
    D_t \big( u^3 - \tfrac{1}{2} u_x^2 \big) + 
    D_x \big( \tfrac{9}{2} u^4 - 6 u u_x^2 + 3 u^2 u_{2x} 
         + \tfrac{1}{2} u_{2x}^2 - u_x u_{3x} \big) = 0,
\end{equation}
which corresponds to Boussinesq's moment of instability.  
For (\ref{PDESystem}), each conservation law has the form
\begin{equation}
  \label{conservationLaw}
  D_t \rho(\vec{u}(x,t),\vec{u}_x(x,t),\dotsc) + 
  D_x J(\vec{u}(x,t),\vec{u}_x(x,t),\dotsc) = 0,
\end{equation}
where $\rho$ is the conserved density and $J$ is the associated flux. 

Algorithms for computing conserved densities and generalized symmetries 
are described in~\cite{Goktas97,GoktasHereman99,Goktas98,Goktas97code,%
WHetalbook08,Heremanetal05,WHandUG99}.  
Our code, 
\texttt{PDE\-Re\-cur\-sion\-Op\-er\-a\-tor.m}~\cite{recursioncode05}, 
uses these algorithms to compute the densities and generalized symmetries 
needed to construct the non-local part of the operator. 
For the benefit of the reader, we present an abbreviated version of these 
algorithms.  

The KdV equation (\ref{kdv}) has conserved densities for any even rank.  
To find the conserved density $\rho$ of rank $R = 6,$ we consider all the 
terms of the form
\begin{equation}
  \label{rankterms}
  D_x^{R - W(u) i} u^i(x,t), \qquad 1 \le i \le R/W(u),
\end{equation}  
where $D_x$ is the total derivative with respect to $x$.  
Hence, since $W(u) = 2$, we have 
\begin{equation}
  D_x^4 u = u_{4x}, \qquad D_x^2 u^2 = 2u_x^2 + 2uu_{2x}, 
    \qquad D_x^0 u^3 = u^3.
\end{equation}
We then remove divergences and divergence equivalent terms 
\cite{Heremanetal05}, 
and take a linear combination (with undetermined coefficients) of the 
remaining terms as the candidate $\rho.$  
Terms are divergence equivalent if and only if they differ by a divergence, 
for instance $uu_{2x}$ and $-u_x^2$ are divergence equivalent because 
$u u_{2x} - (-u_x^2) = D_x(u u_x).$  
Divergences are divergence equivalent to zero, such as $u_{4x} = D_x(u_{3x}).$
Thus, the candidate $\rho$ of rank $R = 6$ is
\begin{equation}
  \rho = c_1 u^3 + c_2 u_x^2. 
\end{equation}
To determine the coefficients $c_i,$ we require that (\ref{conservationLaw}) 
holds on the solutions of (\ref{PDESystem}).  
In other words, we first compute $D_t \rho$ and use (\ref{PDESystem}) 
to remove $\vec{u}_t, \vec{u}_{tx}, $ etc. 

For the KdV equation, 
\begin{equation}
\label{DtrhoKdV}
D_t \rho = 
- ( 18 c_1 u^3 u_x + 3 c_1 u^2 u_{3x} + 12 c_2 u_x^3 + 12 c_2 u u_x u_{2x}
+ 2 c_2 u_x u_{4x} ),
\end{equation}
after $u_t, u_{tx},$ etc.\ have been replaced using (\ref{kdv}).
Then, we require that $D_t \rho$ is a total derivative with respect to $x.$  
To do so, for each component $u_j$ of $\vec{u}$, we apply the Euler operator 
(variational derivative) to $D_t\rho$ and set the result identically equal 
to zero~\cite{Goktas97}.  
The Euler operator for $\vec{u}$ is defined as 
\begin{equation}
\label{euleroperator}
  \Euler{\vec{u}} = \sum_{k=0}^m (-1)^k D_x^k 
  \frac{\partial}{\partial {\vec{u}_{kx}}},
\end{equation}
where $m$ is the highest order needed.
In our scalar example there is only one component ($\vec{u} = u$). 
Using (\ref{DtrhoKdV}), which is of order $m = 4,$
\begin{equation}
  \Euler{u}(D_t \rho) 
    = - 18 (c_1 + 2 c_2) u_x u_{2x}  \equiv 0.
\end{equation}
%
To find the undetermined coefficients, 
we consider all monomials in $u$ and its derivatives as independent, 
giving a linear system for $c_i.$ 
%
For the example, 
$c_1 + 2 c_2 = 0,$ and taking $c_1 = 1$ and $c_2 = -\tfrac{1}{2}$ gives
\begin{equation}
  \rho = u^3 - \tfrac{1}{2} u_x^2,
\end{equation}
which is the conserved density in conservation law (\ref{whitham}).  

%
%
\subsection{Computation of Generalized Symmetries}
\label{subsec:AlgoGenSym}

A generalized symmetry, $\vec{G}(\vec{u}),$ leaves the PDE invariant under 
the replacement $\vec{u} \to \vec{u} + \epsilon \vec{G}$ within order 
$\epsilon$~\cite{Olver93}. 
Hence, $\vec{G}$ must satisfy the linearized equation
\begin{equation}
  \label{symLinEq}
  D_t \vec{G} = \vec{F}'(\vec{u})[\vec{G}],
\end{equation}
where $\vec{F}'(\vec{u})[\vec{G}]$ is the Fr\'echet derivative of 
$\vec{F}$ in the direction of $\vec{G},$  
\begin{equation}
  \label{frechet}
  \vec{F}'(\vec{u})[\vec{G}] = 
    \frac{\partial}{\partial \epsilon}\left. \vec{F}(\vec{u} 
      + \epsilon \vec{G}) \right|_{\epsilon = 0}
    = \sum_{i=0}^m (D_x^i \vec{G}) 
      \frac{\partial \vec{F}}{\partial \vec{u}_{ix}},
\end{equation}
where $m$ is the order of ${\bf F}.$
The KdV equation (\ref{kdv}) has generalized symmetries for any odd rank.  
To find the generalized symmetry of rank $R = 7,$ we again consider the 
terms in (\ref{rankterms}).  
This time we do not remove the divergences or divergence equivalent terms.  
The candidate generalized symmetry is then the linear combination of the 
monomials generated by (\ref{rankterms}).  
For the example, where $W(u) = 2$, 
\begin{equation}
  D_x^5 u = u_{5x}, \qquad D_x^3 u^2 = 6 u_x u_{2x} + 2 u u_{3x}, 
    \qquad D_x u^3 = 3 u^2 u_x,
\end{equation}
so the candidate generalized symmetry of rank $R = 7$ is 
\begin{equation}
  G = c_1 u^2 u_x + c_2 u_x u_{2x} + c_3 u u_{3x} + c_4 u_{5x}.
\end{equation}
The undetermined coefficients are then found by computing (\ref{symLinEq}) 
and using (\ref{PDESystem}) to remove $u_t, u_{tx}, u_{txx},$ etc. 
Thus, continuing with the example we have
\begin{multline}  
\label{systemforKdV}
  2 (2 c_1 - 3 c_2) u_x^2 u_{2x} + 2 (c_1 - 3 c_3) u u_{2x}^2 
    + 2(c_1 - 3 c_3) u u_x u_{3x} + (c_2 - 20 c_4) u_{3x}^2 \\
  + (c_2 + c_3 - 30 c_4) u_{2x} u_{4x} 
  + (c_3 - 10 c_4) u_x u_{5x} \equiv 0.
\end{multline}
%
%
Again, considering all monomials in $u$ and its derivatives as independent 
gives a linear system for $c_i.$ 
%
From (\ref{systemforKdV}),
\begin{equation}
  c_1 = 30 c_4, \qquad c_2 = 20 c_4, \qquad c_3 = 10 c_4.
\end{equation}
Setting $c_4 = 1,$ we find 
\begin{equation}
  G = 30 u^2 u_x + 20 u_x u_{2x} + 10 u u_{3x} + u_{5x},
\end{equation}
which is the fifth-order symmetry $G^{(3)}$ in (\ref{kdvSym}).  

%
%
\section{Algorithm for Computing Recursion Operators}
\label{sec:algoRecursion}

A recursion operator, $\R,$ is a linear integro-differential operator 
which links generalized symmetries~\cite{Olver93},
\begin{equation}
\label{nonsequentialsymmetries}
  \vec{G}^{(j+g)} = \R\vec{G}^{(j)}, \qquad j=1,2,3,\dotsc,
\end{equation}
where $g$ is the gap and $\vec{G}^{(j)}$ is the $j$-th generalized symmetry.  
In many cases, $g = 1$ because the generalized symmetries differ by a common 
rank {\em and}, starting from $\vec{G}^{(1)},$ all higher-order symmetries 
can indeed be consecutively generated with the recursion operator.
However, there are exceptions \cite{DBthesis} where $g = 2$ or $3.$ 
Examples of the latter are given in Sections~\ref{subsec:kkrecursion}, 
\ref{subsec:hirotasatsuma}, \ref{subsec:drinfeld}, and 
Appendix~\ref{sec:softwareUsage}.
Inspection of the ranks of generalized symmetries usually provides a 
hint on how to select the gap.

If $\R$ is a recursion operator for (\ref{PDESystem}), then the Lie 
derivative~\cite{Wang98,Olver93,WHandUG99} of $\R$ is zero, which leads to 
the following defining equation:
\begin{equation}
 \label{defining}
 \frac{\partial \R}{\partial t} + \R'[\vec{F}(\vec{u})] + 
 \R \circ \vec{F}'(\vec{u})-\vec{F}'(\vec{u}) \circ \R = 0, 
\end{equation}
where $\circ$ denotes a composition of operators, $\R'[\vec{F}(\vec{u})]$ 
is the Fr\'echet derivative of $\R$ in the direction of $\vec{F},$ 
\begin{equation}
 \label{frechetofr}
  \R'[\vec{F}(\vec{u})] = 
  \sum_{i=0}^m \left(D_x^i \vec{F}(\vec{u})\right) 
    \frac{\partial \R}{\partial \vec{u}_{ix}},
\end{equation}
and $\vec{F}'(\vec{u})$ is the Fr\'echet derivative operator, 
i.e.\ a $M \times M$ matrix with entries 
\begin{equation}
  \label{frechetFu}
  \vec{F}_{ij}'(\vec{u}) = \sum_{k=0}^{m} \left( 
  \frac{\partial F_i}{\partial (u_j)_{k x}} \right) \, D^{k}_x, 
\end{equation} 
where $m$ is the highest order occurring in the right hand side of 
(\ref{PDESystem}).

In the scalar case, $\vec{F} = F$ and $\vec{u} = u,$ (\ref{frechetFu}) 
simplifies into 
\begin{equation}
  \label{frechetFuscalar}
  F'(u) = \sum_{k=0}^{m} \left( 
  \frac{\partial F}{\partial u_{kx}} \right) \, D^{k}_x.
\end{equation} 
Rather than solving (\ref{defining}), we will construct a candidate recursion 
operator and use (\ref{defining}) to determine the unknown coefficients, 
as shown in the following two steps.
\vskip 5pt
\noindent
\step{Generate the candidate recursion operator}
\vskip 3pt
\noindent
The rank of the recursion operator is determined by the difference in 
ranks of the generalized symmetries the recursion operator actually connects, 
\begin{equation}
  \label{rankRelation}
  \rank\,\R_{ij} = \rank\, \vec{G}_i^{(k+g)} -\rank\, \vec{G}_j^{(k)}, 
  \quad i,j = 1, \ldots, M,
\end{equation}
where $\R$ is an $M\times M$ matrix and $\vec{G}$ has $M$ components.
As before, $g$ is the gap and typically $g = 1$. 
Yet, there are cases where $g = 2$ or $3$.

The recursion operator naturally splits into two pieces~\cite{BaldwinCRM2005},
\begin{equation}
\label{sumR0andR1}
\R = \R_0 + \R_1,
\end{equation}
where $\R_0$ is a (local) differential operator and $\R_1$ is a (non-local)
integral operator.  

The differential operator $\R_0$ is a linear combination of terms 
\begin{equation}
  D_x^{k_0} u_1^{k_1} u_2^{k_2} \dotsb u_M^{k_M} I, 
    \qquad k_0,k_1,\dotsc \in \mathbb{N},
\end{equation}
where the $k_i$ are non-negative integers taken so the monomial has the 
correct rank and the operator $D_x$ has been propagated to the right.  
%
%
For example,  
\begin{equation}
D_x^2 u I 
= D_x (D_x u I) 
= D_x ( u_x I + u D_x)  
= u_{2x} I + 2 u_x D_x + u D_x^2,
\end{equation}
which, after multiplying the terms by undetermined coefficients, leads to
\begin{equation}
\label{R0candidateKdV}
\R_0 = c_1 u_{2x} I + c_2 u_x D_x + c_3 u D_x^2.
\end{equation}
We will assume that the integral operator $\R_1$ is a linear combination 
of terms 
\begin{equation}
\label{r1construction}
  G^{(i)} D_x^{-1} \otimes \Euler{\vec{u}}(\rho^{(j)}),
    \qquad i,j \in \mathbb{N},
\end{equation}
of the correct rank~\cite{Bilge93,Wang98}.  
In (\ref{r1construction}), $\otimes$ is the matrix outer product, and 
$\Euler{\vec{u}}(\rho^{(j)})$ is the cosymmetry 
(Euler operator applied to $\rho^{(j)}).$
To standardize $\R_1$, propagate $D_x$ to the left.
For example, by integration by parts, 
$D_x^{-1} u_x D_x = u_x I - D_x^{-1} u_{2x} I.$

As shown in~\cite{Bilge93}, the integral operator $\R_1$ can also be computed 
as a linear combination of the terms 
\begin{equation}
  G^{(i)} D_x^{-1} \otimes \psi^{(j)}, 
    \qquad i,j \in \mathbb{N},
\end{equation}
of the correct rank, where $\psi^{(j)}$ is the covariant 
(Fr\'echet derivative of $\rho^{(j)}).$
While $G^{(i)} D_x^{-1} \otimes \Euler{\vec{u}}(\rho^{(j)})$ is 
strictly non-local, $G^{(i)} D_x^{-1} \otimes \psi^{(j)}$ contains both 
differential and integral terms.  
Therefore, it is computationally more efficient to build the candidate 
recursion operator using 
$\Euler{\vec{u}}(\rho^{(j)})$ instead of $\psi^{(j)}.$ 
%
Finally, the local and non-local operators are added to obtain a candidate 
recursion operator (\ref{sumR0andR1}).  
\vfill
\newpage
\noindent
\step{Determine the unknown coefficients}
\vskip 3pt
\noindent
To determine the unknown coefficients in the recursion operator,
we substitute the candidate into the defining equation (\ref{defining}).  
After normalizing the form of the terms 
(propagating the $D_x$ through the expression toward the right), 
we group the terms in like powers of $\vec{u},\vec{u}_x,\vec{u}_{xx},\dotsc,
I, D_x, D_x^2, \dotsc,$ and $D_x^{-1}.$  
Requiring that these terms vanish identically gives a linear system for 
the $c_i.$  
Solving this linear system and substituting the coefficients into the 
candidate operator gives the recursion operator for (\ref{PDESystem}).  
%
%
If $c_i = 0$ for all $i,$ then either the gap $g$ is incorrect or 
(\ref{PDESystem}) does not have a recursion operator.  

While the gap $(g)$ is usually $1, 2$ or $3,$ it is not obvious which 
value to take for $g.$
In Sections~\ref{subsec:kkrecursion} and~\ref{subsec:hirotasatsuma}
we give a couple of examples where $g = 2.$  
Starting from $\vec{G}^{(1)},$ the recursion operator then generates the 
higher-order symmetries $\vec{G}^{(3)}, \vec{G}^{(5)}, \ldots.$ 
Analogously, starting from $\vec{G}^{(2)}$ the recursion operator produces 
$\vec{G}^{(4)}, \vec{G}^{(6)},$ and so on.
In Section~\ref{subsec:drinfeld} we show an example where $g = 3.$
Further details on how to select the gap are given in 
Section~\ref{sec:keyAlgorithms}.

%
%
\section{Examples}
\label{sec:examples}

%
%
\subsection{The Korteweg-de Vries Equation}
\label{subsec:kdv}
To illustrate the method, we use the KdV equation (\ref{kdv}). 
Reversing the sign of $t,$
\begin{equation}
  \label{kdvagain}
  u_t = F(u) = 6 u u_x + u_{3x}
\end{equation}
for scalar $u(x,t).$
From (\ref{kdvSym}), the difference in ranks of the generalized symmetries is
\begin{equation}
  \rank\,G^{(3)} - \rank\,G^{(2)} = \rank\,G^{(2)} - \rank\,G^{(1)} = 2.
\end{equation}  
Therefore, we will assume $g=1,$ and build a recursion operator with 
$\rank\,\R = 2.$
Thus, the local operator has the terms $D_x^2$ and $D_x^0 u I = u I$ of 
rank $2.$ 
So, the candidate differential operator is 
\begin{equation}
\label{R0kdv}
  \R_0 = c_1 D_x^2 + c_2 u I.
\end{equation}
Using $\rho^{(1)} = u$ and $G^{(1)} = u_x,$ the non-local operator is 
\begin{equation}
\label{R1kdvoriginal}
  \R_1 = c_3 G^{(1)} D_x^{-1} \Euler{u}(\rho^{(1)}) 
       = c_3 u_x D_x^{-1}\Euler{u}(u) = c_3 u_x D_x^{-1}, 
\end{equation}
where we used $\Euler{u}$ given in (\ref{euleroperator}).
Thus, the candidate recursion operator is 
\begin{equation}  
  \label{candidatekdvR}
  \R = \R_0 + \R_1 = c_1 D_x^2 + c_2 u I + c_3 u_x D_x^{-1}.
\end{equation}
Note that each term in (\ref{candidatekdvR}) indeed has rank 2.

Now, we separately compute the pieces needed to evaluate (\ref{defining}). 
Using (\ref{frechetFuscalar}), 
%
%
we readily compute
\begin{equation}  
  \label{FrechetonF}
  F'(u) = D_x^{3} + 6 u D_x + 6 u_x I.
\end{equation}
Since the candidate recursion operator (\ref{candidatekdvR}) is 
$t$-independent, we have ${\partial \R}/{\partial t} = 0.$ 
Next, using (\ref{frechetofr}) and (\ref{candidatekdvR}) we compute
\begin{equation}
\label{frechetofrkdv}
  \R'[F(u)] = ( 6 c_{2} u u_{x} + c_{2} u_{3 x} ) I
              + ( 6 c_{3} u_{x}^{2} + 6 c_{3} u u_{2 x} 
                  + c_{3} u_{4 x} ) D_x^{-1}.
\end{equation}
Using (\ref{candidatekdvR}) and (\ref{FrechetonF}), we compute
\begin{align}
\label{rafterfprimekdv}
  \R \circ F'(u) 
    & = c_{1} D_x^{5} + ( 6 c_{1} + c_{2} ) u D_x^{3} 
        + ( 18 c_{1} + c_{3} ) u_{x} D_x^{2} \\ 
    \notag & \qquad
    + 6 ( c_{2} u^{2} + 3 c_{1} u_{2x} ) D_x 
    + 6 ( c_{2} u u_{x} + c_{3} u u_{x} + c_{1} u_{3 x} ) I,
\end{align}
and
\begin{align}
\label{fprimeafterrkdv}
     F'(u) \circ \R 
     & = c_{1} D_x^{5} + (6 c_{1} + c_{2} ) u D_x^{3} 
     + ( 6c_{1} + 3c_{2} + c_{3} ) u_{x} D_x^{2} \\ 
     \notag & \qquad
     + 3 ( 2c_{2} u^{2} + c_{2} u_{2 x} + c_{3}u_{2 x} ) D_x 
     + ( 12c_{2} u u_{x} + 6 c_{3} u u_{x} \\ 
     \notag & \qquad
     + c_{2}u_{3 x} + 3c_{3} u_{3 x} ) I 
     + ( 6c_{3} u_{x}^{2} + 6c_{3} u u_{2 x} + c_{3} u_{4 x} ) D_x^{-1}.
\end{align}
%
%
%
Substituting (\ref{frechetofrkdv}), (\ref{rafterfprimekdv}) and 
(\ref{fprimeafterrkdv}) into (\ref{defining}) and grouping like terms, we find 
\begin{equation}
  (4 c_{1} - c_{2}) u_{x} D_x^{2} + (6 c_{1} - c_{2} - c_{3}) u_{2 x} D_x 
    + (2 c_{1}- c_{3}) u_{3 x} I \equiv 0.
\end{equation}
So, $2 c_1 = c_3$ and $c_2 = 2 c_3.$  
Taking $c_3 = 2,$ gives 
\begin{equation}
  \R = D_x^2 + 4u I + 2 u_x D_x^{-1},
\end{equation}
which is indeed the recursion operator (\ref{kdvRecursion}) of the 
KdV equation~\cite{Olver77}.

%
%
\subsection{The Kaup-Kupershmidt Equation}
\label{subsec:kkrecursion}

Consider the Kaup-Kupershmidt (KK) equation~\cite{Wang98,Goktas97},
\begin{equation}
  \label{kk}
  u_t = F(u) = 20 u^2 u_x + 25 u_x u_{2x} + 10 u u_{3x} + u_{5x}.
\end{equation}
To find the dilation symmetry for (\ref{kk}), we require that all the terms 
in (\ref{kk}) have the same rank:
\begin{multline}
  W(u) + W(D_t) = 3 W(u) + W(D_x) = 2 W(u) + 3 W(D_x) \\
    = 2 W(u) + 3 W(D_x) = W(u) + 5 W(D_x).
\end{multline}
If we set $W(D_x) = 1,$ then $W(u) = 2, W(D_t) = 5$ and the rank of 
(\ref{kk}) is $7.$  

Using \texttt{In\-var\-i\-ants\-Sym\-met\-ries.m}~\cite{Goktas97code}, 
we compute the conserved densities 
\begin{equation}
\label{densitiesKK}
\rho^{(1)} \!=\! u, \qquad \rho^{(2)} \!=\! - 8 u^3 + 3u_x^2,
\end{equation}
and the generalized symmetries 
\begin{equation}
\label{symmetriesKK}
G^{(1)} \!=\! u_x, \qquad
G^{(2)} \!=\! F(u) \!=\! 20 u^2 u_x + 25 u_x u_{2x} +10uu_{3x} + u_{5x} 
\end{equation}
of (\ref{kk}).  
%
We do not show $G^{(3)}$ through $G^{(6)}$ explicitly due to length.
From the weights above, the ranks of the first six generalized symmetries are 
\begin{equation}
  \begin{aligned}
   \rank\,G^{(1)} &= 3, & \rank\,G^{(2)} &= 7, & \rank\,G^{(3)} &= 9, \\
   \rank\,G^{(4)} &= 13, & \rank\,G^{(5)} &= 15, & \rank\,G^{(6)} &= 19.
  \end{aligned}
\end{equation}
We assume that $\rank\,{\R} = 6$ and $g = 2,$ since 
$\rank\,G^{(2)} - \rank\,G^{(1)} \neq \rank\,G^{(3)} - \rank\,G^{(2)}$ but 
$\rank\,G^{(3)} - \rank\,G^{(1)} = \rank\,G^{(4)} - \rank\,G^{(2)} = 6.$
%
Thus, taking all terms of the form $D_x^iu^j$ $(i,j\in\mathbb{N})$ such that 
$\rank\,{(D_x^iu^j)} = 6$ gives
\begin{multline}
  \R_0 = c_1D_x^6 + c_2uD_x^4 + c_3u_xD_x^3 
      + ( c_4u^2 + c_5u_{2x})D_x^2  \\
      + (c_6 uu_x + c_7 u_{3x}) D_x + (c_8 u^3 + c_9 u_x^2 
      + c_{10} uu_{2x} + c_{11} u_{4x}) I.
\end{multline}
Using the densities and generalized symmetries above, we compute
\begin{equation}
  G^{(1)} D_x^{-1} \Euler{u}(\rho^{(2)}) = 
    u_x D_x^{-1} \Euler{u}(- 8 u^3 + 3 u_x^2) = 
    - 6 u_x D_x^{-1} (4 u^2 + u_{2x}) 
\end{equation}
and
\begin{equation}
  G^{(2)} D_x^{-1} \Euler{u}(\rho^{(1)}) = 
    F(u) D_x^{-1} \Euler{u}(u) = 
    F(u) D_x^{-1}.
\end{equation}
Thus, the candidate non-local operator is
\begin{equation}
  \R_1 = c_{12} u_x D_x^{-1} ( 4 u^2 + u_{2x} ) I +  
    c_{13} \left( 
      20 u^2 u_x + 25u_x u_{2x} + 10 uu_{3x} + u_{5x} \right) D_x^{-1}. 
\end{equation}
Substituting $\R \!=\! \R_0 \!+\! \R_1$ into (\ref{defining}) gives $49$ 
linear equations for $c_i.$  
Solving yields
\begin{equation}
  \begin{gathered}
c_{1} = - \frac{1}{2} c_{12}, \;\;\; 
c_{2} = -6 c_{12}, \;\;\;
c_{3} = c_{5} = -18 c_{12}, \;\;\;
c_{4} = -16 c_{12},\\
c_{6} = -\frac{69}{2} c_{12}, \;\;\; 
c_{7} = -\frac{49}{2} c_{12}, \;\;\;
c_{8} = -\frac{35}{2} c_{12}, \;\;\;
c_{9} = -\frac{13}{2} c_{12}, \\
c_{10} = -60 c_{12}, \;\;\; 
c_{11} = -41 c_{12}, \;\;\;
c_{13} = -c_{12}, \;\;\;
  \end{gathered}
\end{equation}
where $c_{12}$ is arbitrary.  
Setting $c_{12} = -2,$ we obtain
\begin{multline}
  \R = D_x^6 + 12 u D_x^4 + 36 u_x D_x^3 
    + \left( 36 u^2 + 49 u_{2x} \right) D_x^2 \\
    + 5 \left( 24 uu_x + 7 u_{3x} \right) D_x 
    + \left( 32 u^3  + 69 u_x^2 + 82 uu_{2x} + 13 u_{4x} \right) I \\
    + 2 u_x D_x^{-1} \left( 4 u^2 + u_{2x} \right) I + 2 F(u) D_x^{-1},
\end{multline}
which was computed in~\cite{Wang98} as the composition of the cosymplectic 
and symplectic operators of (\ref{kk}).

Since $g = 2$ the symmetries are not generated sequentially via 
(\ref{sequentialsymmetries}).
Actually, $G^{(1)}$ and $G^{(2)}$ in (\ref{symmetriesKK}) are the 
``seeds" (or roots) and one must use (\ref{nonsequentialsymmetries}).
Indeed, from $G^{(1)}$ one obtains $G^{(3)} = \R G^{(1)}, 
G^{(5)} = \R G^{(3)},$ and so on.
From $G^{(2)},$ upon repeated application of $\R,$ one obtains 
$G^{(4)}, G^{(6)},$ etc.
Thus, using the recursion operator one can generate an infinity of 
generalized symmetries, confirming that (\ref{kk}) is completely integrable.  

%
%
\subsection{The Hirota-Satsuma System}
\label{subsec:hirotasatsuma}

Consider the system of coupled KdV equations due to Hirota and 
Satsuma~\cite{Ablowitz91},
\begin{equation}
  \label{hirota}
  \begin{aligned}
    u_t & = F_1(\vec{u}) = 3 u u_x - 2 v v_x + \frac{1}{2} u_{3x}, \\
    v_t & = F_2(\vec{u}) = - 3 u v_x - v_{3x},
  \end{aligned} 
\end{equation} 
which model shallow water waves.
Solving the equations for the weights, 
\begin{equation}
  \begin{cases}
    W(u) + W(D_t) = 2W(u) + 1 = W(u) + 3 = 2 W(v) + 1, \\
    W(v) + W(D_t) = W(u) + W(v) + 1 = W(v) + 3,
  \end{cases}
\end{equation}
yields $W(u) = W(v) = 2$ and $W(D_t) = 3.$ 

The first few conserved densities and generalized symmetries computed with
\texttt{In\-var\-i\-ants\-Sym\-met\-ries.m}~\cite{Goktas97code} are 
\begin{equation}
\label{densitiessymmetrieshirota}
  \begin{gathered}
  \rho^{(1)} = u, \qquad \rho^{(2)} = 3 u^2 - 2 v^2, \\
  \vec{G}^{(1)} = 
    \begin{pmatrix} 
    u_x \\ v_x \end{pmatrix}, \qquad 
  \vec{G}^{(2)} = 
    \begin{pmatrix}  
      F_1(\vec{u}) \\ 
      F_2(\vec{u}) 
    \end{pmatrix}
    = \begin{pmatrix}  
      3 u u_x - 2 v v_x + \tfrac{1}{2} u_{3x}\\ 
      - 3 u v_x - v_{3x} 
    \end{pmatrix}.
  \end{gathered}
\end{equation}
$G^{(3)}$ and $G^{(4)}$ are not shown explicitly due to length.
Based on the above weights, 
\begin{equation}
  \rank\,\rho^{(1)} = 2,\qquad \rank\,\rho^{(2)} = 4, \\
\end{equation}
and 
\begin{equation}
\label{ranksymmetrieshirota}
  \begin{gathered}
  \rank\,\vec{G}^{(1)} = 
  \begin{pmatrix} 
  3 \\ 3 
  \end{pmatrix}, \qquad
  \rank\,\vec{G}^{(2)} = 
  \begin{pmatrix} 
  5 \\ 5 
  \end{pmatrix}, \\
  \rank\,\vec{G}^{(3)} = 
  \begin{pmatrix} 
  7 \\ 7 
  \end{pmatrix}, \qquad
  \rank\,\vec{G}^{(4)} = 
  \begin{pmatrix} 
  9 \\ 9 
  \end{pmatrix}. 
  \end{gathered}
\end{equation}
We first set $g = 1,$ so that $\rank\,\R_{ij} = 2, \,i,j = 1,2.$  
If indeed the generalized symmetries were linked consecutively, then 
\begin{equation}
  \R_0 = 
    \left( 
    \begin{array}{cc} 
    c_1 D_x^2 + c_2 u I + c_3 v I \quad & c_4 D_x^2 + c_5 u I + c_6 v I\\
    c_7 D_x^2 + c_8 u I + c_9 v I \quad 
    & c_{10} D_x^2 + c_{11} u I + c_{12} v I
    \end{array} 
    \right).
\end{equation}
Using (\ref{densitiessymmetrieshirota}), we have 
\begin{align}
  \R_1 & = c_{13} \vec{G}^{(1)} D_x^{-1} \otimes \Euler{\vec{u}}( \rho^{(1)}) 
     = c_{13} 
      \begin{pmatrix} u_x \\ v_x \end{pmatrix} 
      D_x^{-1} \otimes 
      \begin{pmatrix} 
        \Euler{u}( \rho^{(1)}) & \Euler{v}( \rho^{(1)}) 
      \end{pmatrix} \\
   & = c_{13} \begin{pmatrix} u_x \\ v_x \end{pmatrix} 
      D_x^{-1} \otimes 
      \begin{pmatrix} I & 0  
      \end{pmatrix} 
    = c_{13}
      \begin{pmatrix}  
        u_x D_x^{-1} \quad & 0 \\ 
        v_x D_x^{-1} \quad & 0
      \end{pmatrix}.
\end{align}
Substituting $\R = \R_0 + \R_1 $ into (\ref{defining}), 
we find $c_1 = \dotsb = c_{13} = 0.$ 
Thus, the choice $g \!=\! 1$ appears to be {\it incorrect}.
Noting that the ranks of the symmetries in (\ref{ranksymmetrieshirota}) 
differ by 2, we repeat the process with $g \!=\! 2,$ so that 
$\rank\,\R_{ij} = 4,\,i,j = 1,2.$
%
%
Thus,
\begin{equation}
\label{totalrecursionoperator}
    \R = 
      \begin{pmatrix} 
        ({\R_0})_{11} & ({\R_0})_{12} \\
        ({\R_0})_{21} & ({\R_0})_{22}
      \end{pmatrix} +
      c_{41}\vec{G}^{(1)} D_x^{-1} \otimes \Euler{\vec{u}}(\rho^{(2)})
      + c_{42}\vec{G}^{(2)} D_x^{-1} \otimes \Euler{\vec{u}}(\rho^{(1)}),
\end{equation}
where $({\R_0})_{ij},\,i,j = 1,2,$ are linear combinations (with different 
undetermined coefficients) of 
$\{ D_x^4, u D_x^2, v D_x^2, u_x D_x, v_x D_x, u^2, uv, v^2, u_{2x}, v_{2x}\}.$
For instance, 
\begin{multline}
  ({\R_0})_{12} = c_{11} D_x^4 + (c_{12} u + c_{13} v ) D_x^2 
  + ( c_{14} u_x + c_{15} v_x ) D_x \\ 
  + ( c_{16} u^2  + c_{17} uv + c_{18} v^2 + c_{19} u_{2x} + c_{20} v_{2x} ) I.
\end{multline}
Using (\ref{densitiessymmetrieshirota}), the first term of $\R_1$ in 
(\ref{totalrecursionoperator}) is
\begin{align*}
 \R^{(1)}_1 = 
 c_{41}\vec{G}^{(1)} D_x^{-1} \otimes \Euler{\vec{u}}(\rho^{(2)}) & = 
 c_{41}
 \begin{pmatrix} 
       u_x \\ v_x \end{pmatrix} 
       D_x^{-1} \otimes 
       \begin{pmatrix} 
       6 u I & -4 v I
       \end{pmatrix} \\
  & =
  c_{41}
    \begin{pmatrix}
      3 u_x D_x^{-1} u I \quad & -2 u_x D_x^{-1} v I \\
      3 v_x D_x^{-1} u I \quad & -2 v_x D_x^{-1} v I \\
    \end{pmatrix}.
\end{align*}
The second term of $\R_1$ in (\ref{totalrecursionoperator}) is
\begin{align*}
 \R^{(2)}_1 =  
 c_{42}\vec{G}^{(2)} D_x^{-1} \otimes \Euler{\vec{u}}(\rho^{(1)}) & = 
    c_{42} \begin{pmatrix} F_1(\vec{u}) \\ F_2(\vec{u}) \end{pmatrix} 
      D_x^{-1} \otimes \begin{pmatrix} I & 0 \end{pmatrix} \\
 & = c_{42} \begin{pmatrix} 
      F_1(\vec{u})D_x^{-1} \quad & 0 \\
      F_2(\vec{u})D_x^{-1} \quad & 0 \\
    \end{pmatrix}.
\end{align*}
Substituting the form of $\R = \R_0 + \R_1 = \R_0 + \R^{(1)}_1 + \R^{(2)}_1$ 
into (\ref{defining}), the linear system for $c_i$ has a non-trivial solution. 
Solving the linear system, we finally obtain
\begin{equation}
  \label{hsRec}
  \R = 
    \begin{pmatrix} 
      (\R)_{11} \quad & (\R)_{12} \\
      (\R)_{21} \quad & (\R)_{22}
    \end{pmatrix},
\end{equation}
with 
\begin{align*}
(\R)_{11} & =
  D_x^4 + 8u D_x^2 + 12u_{x} D_x 
  + 8 \left( 2 u^{2} + u_{2x} - \frac{2}{3} v^{2} \right) I \\
    & \qquad + 4u_{x} D_x^{-1} u I 
    + 2 \left( 6uu_{x} + u_{3x} - 4vv_{x} \right) D_x^{-1}, \notag \\
(\R)_{12} & =
  -\frac{20}{3}vD_x^2 - \frac{16}{3}v_{x}D_x 
    - \frac{4}{3} \left( 4 uv + v_{2 x} \right) I 
    - \frac{8}{3}u_{x}D_x^{-1}v I, \\
(\R)_{21} & =
  -10v_{x}D_x - 12v_{2 x}I + 4v_{x}D_x^{-1}uI
    - 4 \left( 3uv_{x} + v_{3 x} \right) D_x^{-1}, \\
(\R)_{22} & =
  - 4D_x^4 - 16uD_x^2 - 8u_{x}D_x
    - \frac{16}{3}v^{2}I - \frac{8}{3}v_{x}D_x^{-1}vI .
\end{align*}
The above recursion operator was computed in~\cite{Wang98} as the 
composition of the cosymplectic and symplectic operators of (\ref{hirota}).

In agreement with $g = 2$, there are two seeds. 
Using (\ref{nonsequentialsymmetries}) and starting from $\vec{G}^{(1)}$ 
in (\ref{densitiessymmetrieshirota}), the recursion operator (\ref{hsRec}) 
generates the infinite sequence of generalized symmetries with odd labels. 
Starting from $\vec{G}^{(2)},$ the recursion operator (\ref{hsRec}) 
generates the infinite sequence of generalized symmetries with even labels.
The existence of a recursion operator confirms that (\ref{hirota}) is 
completely integrable.  

%
%
\section{Key Algorithms}
\label{sec:keyAlgorithms}

In this section, we present details of the algorithm. 
%
%
To illustrate the key algorithms in Sections~\ref{subsec:candidateOperator}
and~\ref{subsec:coefficients}, we will use the dispersiveless long wave 
system~\cite{Ablowitz91},
\begin{equation}
  \label{dlw}
  \begin{aligned}
    u_t & = F_1(\vec{u}) = u v_x + u_x v, \\
    v_t & = F_2(\vec{u}) = u_x + v v_x,
  \end{aligned}
\end{equation}
which is used in applications involving shallow water waves.  

%
%
\subsection{Integro-Differential Operators}
\label{subsec:operatorAlgebra}

Recursion operators are non-commutative by nature and certain rules must 
be used to simplify expressions involving integro-differential operators.
While the multiplication of differential and integral operators is 
completely described by 
\begin{equation}
  D_x^i D_x^j = D_x^{i+j}, \qquad i,j\in\mathbb{Z},
\end{equation}
the propagation of a differential operator through an expression is trickier.  
To propagate the differential operator to the right, we use Leibniz' rule
\begin{equation}
  \label{propRight}
  D_x^n Q = \sum_{k=0}^n \binom{n}{k} Q^{(k)} D_x^{n-k}, 
    \qquad n \in\ \mathbb{N},
\end{equation}
where $Q$ is an expression and $Q^{(k)}$ is the $k$-th derivative with 
respect to $x$ of $Q.$  
Unlike the finite series for a differential operator, Leibniz' rule for an 
inverse differential operator is 
\begin{equation}
  \label{inverseLiebniz}
  D_x^{-1} Q = Q D_x^{-1} - Q' D_x^{-2} + Q''D_x^{-3} - \dotsb 
    = \sum_{k=0}^\infty (-1)^kQ^{(k)}D_x^{-k-1}.
\end{equation}
Therefore, rather than dealing with an infinite series, 
we only use Leibniz' rule for the inverse differential operator 
when there is a differential operator to the right of the inverse operator.
In such cases we use 
\begin{equation}
\label{propLeftbasic}
  D_x^{-1} Q D_x^n = QD_x^{n-1} - D_x^{-1} Q' D_x^{n-1}.
\end{equation}
Repeated application of (\ref{propLeftbasic}) yields
\begin{equation}
  \label{propLeft}
    D_x^{-1} Q D_x^n = \sum_{k=0}^{n-1} (-1)^kQ^{(k)}D_x^{n-k-1} + 
      (-1)^n D_x^{-1} Q^{(n)}I.
\end{equation}
By using these identities, 
all the terms are either of the form ${\tilde{P}} D_x^n$ 
or ${\tilde{P}} D_x^{-1} {\tilde{Q}} I,$ where ${\tilde{P}}$ and ${\tilde{Q}}$ 
are polynomials in $\vec{u}$ and its $x$ derivatives.  

%
%
\subsection{Algorithm for Building the Candidate Recursion Operator}
\label{subsec:candidateOperator}
%
\vskip 1pt
\noindent
\step{Find the dilation symmetry}
\vskip 3pt
\noindent
The dilation symmetry is found by requiring that each equation in 
(\ref{PDESystem}) is uniform in rank, i.e.\ every monomial in that equation 
has the same rank.  
If (\ref{PDESystem}) is not uniform in rank we use a trick.
In that case, we multiply those terms that are not uniform in rank by 
auxiliary parameters ($\alpha, \beta,\ldots )$ with weights.  
Once the computations are finished we set the auxiliary parameters equal 
to one.  

Since the linear system for the weights is always underdetermined, 
we set $W(D_x) = 1$ and this (typically) fixes the values for the 
remaining weights.

For the example under consideration, (\ref{dlw}), we have the linear system
\begin{equation}
  \begin{gathered}
    W(u) + W(D_t) = W(u) + W(v) + 1 = W(u) + W(v) + 1, \\
    W(v) + W(D_t) = W(u) + 1 = 2 W(v) + 1.
  \end{gathered}
\end{equation}
Thus, $W(v) = \frac{1}{2} W(u), 
W(D_t) = \frac{1}{2} W(u) + 1,$ provided $W(D_x) = 1.$  
If we select $W(u) = 2,$ then the scaling symmetry for (\ref{dlw}) becomes 
\begin{equation}
  (t,x,u,v) \to (\lambda^{-2} t, \lambda^{-1} x, \lambda^2 u, \lambda v).
\end{equation}
In the code \texttt{PDE\-Re\-cur\-sion\-Op\-er\-a\-tor.m}, the user can 
set the values of weights with \verb|WeightRules -> {weight[u] -> 2}|.
%
\vskip 5pt
\noindent
\step{Determine the rank of the recursion operator}
\vskip 3pt
\noindent
Since the gap $g$ cannot be determined {\em a priori}, we assume $g = 1.$ 
Should this choice not lead to a result, one could set $g = 2$ or $3.$
In the code, the user can set the \verb|Gap| to any positive integer value 
(see Appendix~\ref{sec:softwareUsage}).

To determine the rank of the recursion operator, we compute the first 
$g+1$ generalized symmetries and then use 
\begin{equation}
  \label{impRankRelation}
  \rank\,\R_{ij} = \rank\,\vec{G}_i^{(k+g)} -\rank\,\vec{G}_j^{(k)}, 
  \quad i,j = 1, 2, 
\end{equation}
to determine the rank of $\R.$  
Hence, the rank of the recursion operator $\R$ can be represented 
(in matrix form) as follows
\begin{equation}
\label{rankRgeneral}
  \rank\,\R = 
\begin{pmatrix} 
\rank\,\R_{11}\quad & \rank\,\R_{12} \\ 
\rank\,\R_{21}\quad & \rank\,\R_{22} 
\end{pmatrix}.
\end{equation}
In exceptional cases, the rank of the recursion operator might be lower 
(or higher) than computed by (\ref{impRankRelation}). 
In the code, the user has some additional control over the rank of the 
recursion operator.
For example, in an attempt to find a simpler recursion operator, the rank of 
the recursion operator can be shifted down by one by setting
\verb|RankShift -> -1|.
Similarly, to increase the rank of the recursion operator one can set 
\verb|RankShift -> 1| (see Appendix~\ref{sec:softwareUsage}).

For (\ref{dlw}), the first two generalized symmetries and their ranks are 
\begin{gather}
  \vec{G}^{(1)} = \begin{pmatrix} u_x \\ v_x \end{pmatrix}, \qquad 
    \rank\,\vec{G}^{(1)} = 
   \begin{pmatrix} 3 \\ 2 
   \end{pmatrix}, \\
  \vec{G}^{(2)} = 
   \begin{pmatrix} uv_x + vu_x \\ u_x + vv_x 
   \end{pmatrix}, 
   \qquad \rank\,\vec{G}^{(2)} = 
   \begin{pmatrix} 4 \\ 3 
   \end{pmatrix}. 
\end{gather}
Then, using (\ref{impRankRelation}) and (\ref{rankRgeneral}), 
\begin{equation}
\label{rankRdlw}
  \rank\,\R = 
  \begin{pmatrix} 
  1 \; & 2 \\ 
  0 \; & 1 
  \end{pmatrix}.
\end{equation}
%
\vskip 2pt
\noindent
\step{Generate the (local) differential operator $\R_0$}
\vskip 3pt
\noindent
Given the rank of the recursion operator, we take a linear combination of 
\begin{equation}
  D_x^{k_0} u_1^{k_1} u_2^{k_2} \dotsb u_M^{k_M} 
    \alpha^{k_{M+1}} \beta^{k_{M+2}}\dotsb, 
    \qquad k_0,k_1,\dotsc \in \mathbb{N},
\end{equation}
where the $k_i$ are taken so the monomial has the correct rank, 
the operator $D_x$ has been propagated to the right, and
$\alpha,\beta,\dotsc$ are the weighted parameters from Step~$1$ (if present).

For (\ref{dlw}), the (local) differential operator is
\begin{equation}
  \R_0 =  
    \begin{pmatrix}
      c_1 D_x + c_2 v I 
       \quad & c_3 D_x^2 + c_4 u I + c_5 v D_x + c_6 v^2 I + c_7 v_x I \\
      c_8 I 
       \quad & c_9 D_x + c_{10} v I
    \end{pmatrix}.
\end{equation}
%
\vskip 2pt
\noindent
\step{Generate the (non-local) integral operator $\R_1$}
\vskip 3pt
\noindent
Since the integral operator involves the outer product of generalized 
symmetries and cosymmetries, we compute the conserved densities up to 
\begin{equation}
  \label{rankOfDensity}
  \max_{i,j}\{ \rank\,\R_{ij} - \rank\,(\vec{G}^{(1)})_i + W(u_j) + W(D_x) \}, 
  \quad i,j = 1,\ldots, M.
\end{equation}
We add $W(u_j)$ in (\ref{rankOfDensity}) because the Euler operator 
$\Euler{u_j}$ decreases the weight of the conserved density by the weight of 
$u_j.$  
In most cases, we take a linear combination of the terms 
\begin{equation}
  \label{nonlocalterm}
  G^{(i)} D_x^{-1} \otimes \Euler{\vec{u}}(\rho^{(j)}), 
    \qquad i,j \in \mathbb{N},
\end{equation}
of the correct rank as the candidate non-local operator.  
However, there are cases in which we must take a linear combination of the 
monomials in {\em each term} of type (\ref{nonlocalterm}) with
different coefficients.

For (\ref{dlw}), we only need the cosymmetry of density $\rho^{(1)} = v,$
\begin{equation}
    \Euler{\vec{u}}( \rho^{(1)} ) = \begin{pmatrix} 0 \; & I \end{pmatrix}.
\end{equation}
Hence, 
\begin{equation}
  G^{(1)} D_x^{-1} \otimes \Euler{\vec{u}}(\rho^{(1)}) = 
    \begin{pmatrix} 
      0 \quad & u_x D_x^{-1} \\ 
      0 \quad & v_x D_x^{-1}
    \end{pmatrix}.
\end{equation}
Thus, the (non-local) integral operator is 
\begin{equation}
\label{R1dlw}
  \R_1 = 
    \begin{pmatrix} 
      0 \quad & c_{11} u_x D_x^{-1} \\
      0 \quad & c_{12} v_x D_x^{-1}
    \end{pmatrix}.
\end{equation}
\vskip 2pt
\noindent
\step{Add the local and the non-local operators to form $\R$}
\vskip 3pt
\noindent
The candidate recursion operator is 
\begin{equation}
  \R = \R_0 + \R_1.
\end{equation}
So, the candidate recursion operator for (\ref{dlw}) is 
\begin{equation}
  \R = 
    \begin{pmatrix}
      c_1 D_x + c_2 v I \quad & c_3 D_x^2 + c_4 u I 
        + c_5 v D_x + c_6 v^2 I + c_7 v_x I + c_{11} u_x D_x^{-1} \\
      c_8 I \quad & c_9 D_x + c_{10} v I + c_{12} v_x D_x^{-1}
    \end{pmatrix}.
\end{equation}
%
%
%
\subsection{Algorithm for Determining the Unknown Coefficients}
\label{subsec:coefficients}
\vskip 1pt
\noindent
\step{Compute the terms in the defining equation (\ref{defining})}
\vskip 2pt
\noindent
\substep{Compute $\R_t = \frac{\partial \R}{\partial t}$}
\vskip 2pt
\noindent
The computation of $\R_t$ is easy.  
Since the candidate recursion operator is $t$-independent one has 
$\R_t = \vec{0}.$
\vskip 3pt
\noindent
\substep{Compute \mbox{$\R'[\vec{F}(\vec{u})]$}}
\vskip 2pt
\noindent
The Fr\'echet derivative of $\R$ in the direction of $\vec{F}(\vec{u})$ is 
given in (\ref{frechetofr}).
%
%
Unlike the Fr\'echet derivative (\ref{frechet}) of $\vec{F}(\vec{u})$ in the 
direction of $\vec{G}$ (used in the computation of generalized symmetries), 
$\R$ and $\vec{F}(\vec{u})$ in (\ref{frechetofr}) are operators.

Applied to the example (\ref{dlw}) with 2 components,
\vspace{-2mm}
\noindent
\begin{equation}
  \R'[\vec{F}(\vec{u})] = 
  \begin{pmatrix} 
  (\R'[\vec{F}(\vec{u})])_{11} \quad & (\R'[\vec{F}(\vec{u})])_{12} \\
  (\R'[\vec{F}(\vec{u})])_{21} \quad & (\R'[\vec{F}(\vec{u})])_{22}
  \end{pmatrix},
\end{equation}
\vspace{-2mm}
\noindent
where 
\vspace{-2mm}
\noindent
\begin{equation}
  (\R'[\vec{F}(\vec{u})])_{ij} = 
  \sum_{k=0}^m \left(D_x^k \vec{F}(\vec{u})\right) 
    \frac{\partial (\R)_{ij}}{\partial \vec{u}_{kx}}, \quad i,j = 1, 2.
\end{equation}
\vspace{-2mm}
\noindent
Explicitly,
\vspace{-2mm}
\noindent
\begin{align}
  (\R'[\vec{F}(\vec{u})])_{11} & = c_2 \left( u_{x} + vv_{x}\right) I, \\
  (\R'[\vec{F}(\vec{u})])_{21} & = 0, \\
  (\R'[\vec{F}(\vec{u})])_{12} & = \notag
   + c_5 \left( u_{x} + vv_{x} \right) D_x 
   + \Big( c_4uv_{x} + ( 2c_6+ c_4) vu_{x} + 2c_6v^{2}v_{x}  \\ & 
    + c_7vv_{2x} + c_7v_{x}^{2} + c_7u_{2 x} \Big) I 
    + c_{11}( 2u_{x}v_{x} + uv_{2 x} + vu_{2 x})D_x^{-1}, \\
  (\R'[\vec{F}(\vec{u})])_{22}  & = 
  c_{10} \left( u_{x} + vv_{x} \right) I 
  + c_{12} \left( u_{2 x} + vv_{2 x} + v_{x}^{2} \right) D_x^{-1}.
\end{align}
\vskip 0.01pt
\noindent
\substep{Compute $\vec{F}'(\vec{u})$}
\vskip 2pt
\noindent
Use formula (\ref{frechetFu}) to compute $\vec{F}'(\vec{u}).$
Continuing with example (\ref{dlw}), 
\begin{equation}
  \vec{F}'(\vec{u}) = 
    \begin{pmatrix}
      v D_x + v_{x}I \quad & u D_x + u_{x} I \\
      D_x \quad & v D_x + v_{x} I
    \end{pmatrix}.
\end{equation}
\vskip 0.01pt
\noindent
\substep{Compose $\R$ and $\vec{F}'(\vec{u})$}
\vskip 2pt
\noindent
The composition of the $M\times M$ matrices $\R$ and $\vec{F}'(\vec{u})$ 
is an order preserving inner product of the two matrices.  
For example (\ref{dlw}), 
\begin{equation}
 \R \circ \vec{F}'(\vec{u}) = 
 \begin{pmatrix}
 (\R \circ \vec{F}'(\vec{u}))_{11} \quad & (\R \circ \vec{F}'(\vec{u}))_{12} \\
 (\R \circ \vec{F}'(\vec{u}))_{21} \quad & (\R \circ \vec{F}'(\vec{u}))_{22}
 \end{pmatrix},
\end{equation}
with 
\begin{align}
  (\R \circ \vec{F}'(\vec{u}))_{11} & =  \notag
    c_{3}D_x^{3} + (c_{1} + c_{5}) v D_x^{2} 
    + \big( 2c_{1}v_{x} + c_{2}v^{2} + c_{4}u + c_{6}v^{2}  
    + c_{7}v_{x} \big) D_x
    \\ &  
    + \left( c_{1}v_{2 x} + c_{2}vv_{x} + c_{11}u_{x} \right) I, \\
  (\R \circ \vec{F}'(\vec{u}))_{12} & = \notag 
    c_{3}v D_x^{3} 
    + \left( c_{1}u + 3c_{3}v_{x} + c_{5}v^{2} \right) D_x^{2}  
    + \big( 2c_{1}u_{x} + c_{2}uv + 3c_{3}v_{2 x} \\ \notag &      
    + c_{4}uv + 2c_{5}vv_{x} + c_{6}v^{3} + c_{7}vv_{x} \big) D_x 
    + \big( c_{1}u_{2 x} + c_{2}vu_{x} + c_{3}v_{3 x} \\ &  
    + c_{4}uv_{x} + c_{5}vv_{2 x} + c_{6}v^{2}v_{x} + c_{7}v_{x}^{2} 
    + c_{11}vu_{x} \big) I, \\
  (\R \circ \vec{F}'(\vec{u}))_{21} & = 
    c_{9}D_x^{2} 
  + \left( c_{8} + c_{10} \right) vD_x 
  + \left( c_{8} + c_{12} \right) v_{x}I, \\
  (\R \circ \vec{F}'(\vec{u}))_{22} & = \notag
    c_{9}vD_x^{2} + \left( c_{8}u + 2c_{9}v_{x} + c_{10}v^{2} \right) D_x 
    + \big( c_{8}u_{x} + c_{9}v_{2 x} \\  & 
    + c_{10}vv_{x} + c_{12}vv_{x} \big) I.
\end{align}
\vspace{-1mm}
\noindent
Similarly, 
\begin{equation}
 \vec{F}'(\vec{u}) \circ \R = 
 \begin{pmatrix}
 (\vec{F}'(\vec{u}) \circ \R)_{11} \quad & (\vec{F}'(\vec{u}) \circ \R)_{12} \\
 (\vec{F}'(\vec{u}) \circ \R)_{21} \quad & (\vec{F}'(\vec{u}) \circ \R)_{22}
 \end{pmatrix},
\end{equation}
with
\begin{align}
  (\vec{F}'(\vec{u}) \circ \R)_{11} & = 
    c_{1}vD_x^{2} + \left( c_{1}v_{x} + c_{2}v^{2} + c_{8}u)D_x 
    + ( 2c_{2}vv_{x} + c_{8}u_{x} \right) I, \\
  (\vec{F}'(\vec{u}) \circ \R)_{12} & = \notag
    c_{3}vD_x^{3} + \left( c_{3}v_{x} + c_{5}v^{2} + c_{9}u \right) D_x^{2} 
    + \big( c_{4}uv + 2c_{5}vv_{x} + c_{6}v^{3} \\ \notag & 
    + c_{7}vv_{x} + c_{9}u_{x} + c_{10}uv \big) D_x 
    + \big( c_{4}uv_{x} + c_{4}vu_{x} + 3c_{6}v^{2}v_{x} \\ \notag & 
    + c_{7}v_{x}^{2} + c_{7}vv_{2 x} + c_{10}uv_{x} + c_{10}vu_{x} 
    + c_{11}vu_{x} + c_{12}uv_{x} \big) I \\  &
    + \left( c_{11}u_{x}v_{x} + c_{11}vu_{2 x} + c_{12}uv_{2 x} 
    + c_{12}u_{x}v_{x}\right) D_x^{-1}, \\
  (\vec{F}'(\vec{u}) \circ \R)_{21} & =   
    c_{1}D_x^{2} + \left( c_{2} + c_{8} \right) v D_x 
   + \left( c_{2} + c_{8} \right) v_{x} I, \\
  (\vec{F}'(\vec{u}) \circ \R)_{22} & = \notag 
    c_{3}D_x^{3} + \left( c_{5} + c_{9} \right) v D_x^{2} 
    + \big( c_{4}u + c_{5}v_{x} + c_{6}v^{2} + c_{7}v_{x} \\ \notag & 
    + c_{9}v_{x} + c_{10}v^{2} \big) D_x 
    + \big( c_{4}u_{x} + 2c_{6}vv_{x} + c_{7}v_{2 x} + 2c_{10}vv_{x}
    \\ &  + c_{11}u_{x} + c_{12}vv_{x} \big) I 
    + \left( c_{11}u_{2 x} + c_{12}vv_{2 x} + c_{12}v_{x}^{2} \right) D_x^{-1}.
\end{align}
\vskip 0.1pt
\noindent
\substep{Sum the terms in the defining equation}
\vskip 2pt
\noindent
For (\ref{dlw}), summing the terms in the defining equation (\ref{defining}), 
we find 
\vspace{-2mm}
\noindent
\begin{multline}
  c_{3}D_x^{3} + c_{5}vD_x^{2} + (c_{1} + c_{7})v_{x}D_x 
  + \left( c_{4} - c_{8} \right) uD_x + c_{6}v^{2}D_x \\
  + \left( c_{2} - c_{8} + c_{11} \right) u_{x} I 
  + c_{1}v_{2 x}I \equiv 0,
\end{multline}
\vspace{-2mm}
\noindent
\begin{multline}
  \left( c_{1} - c_{9} \right) uD_x^{2} + 2c_{3}v_{x}D_x^{2} 
  + \left( c_{2} - c_{10} \right) uvD_x 
  + \left( c_{5} - c_{9} + 2c_1 \right) u_{x}D_x \\
  + c_{5}vv_{x}D_x + 3c_{3}v_{2 x}D_x 
  - \left( c_{1} - c_{7} \right) u_{2 x}I 
  + \left( c_{4} - c_{10} - c_{12} \right) uv_{x} I \\
  + \left( c_{2} + 2c_{6} - c_{10} \right) vu_{x} I 
  + c_{5}vv_{2 x}I + c_{7}v_{x}^{2}I + c_{3}v_{3 x}I   \\
  + \left( c_{11} - c_{12} \right) uv_{2 x}D_x^{-1} 
  + \left( c_{11} - c_{12} \right) u_{x}v_{x}D_x^{-1} \equiv 0,
\end{multline}
\vspace{-2mm}
\noindent
\begin{equation}
  \left( c_{1} - c_{9} \right) D_x^{2} 
  + \left( c_{2} - c_{10} \right) vD_x 
  + \left( c_{2} - c_{12} \right) v_{x}I \equiv 0,
\end{equation}
\vspace{-2mm}
\noindent
\begin{multline}
  c_{3}D_x^{3} + c_{5}vD_x^{2} + \left( c_{4} - c_{8} \right) uD_x 
  + \left( c_{5} + c_{7} - c_{9} \right) v_{x}D_x  \\
  + c_{6}v^{2}D_x + \left( c_{4} - c_{8} - c_{10} + c_{11}\right) u_{x} I 
  + 2c_{6}vv_{x} I \\
  + \left( c_{7} - c_{9} \right) v_{2 x} I 
  + \left( c_{11} - c_{12} \right) u_{2 x}D_x^{-1} \equiv 0.
\end{multline}
%
\vskip 0.1pt
\noindent
\step{Extract the linear system for the undetermined coefficients}
\vskip 2pt
\noindent
Group the terms in like powers of $\vec{u}, \vec{u}_x, \vec{u}_{xx},
\dotsc, I,D_x,D_x^2,\dotsc$ and $D_x^{-1}.$  
Then, grouping like terms and setting their coefficients equal to zero yields
a linear system for the undetermined coefficients.  
For (\ref{dlw}), we obtain 
\begin{equation}
  \begin{gathered}
    c_{1} = 0, \quad c_{3} = 0, \quad c_{5} = 0, \quad c_{6} = 0, 
    \quad c_{7} = 0, \quad c_{4} - c_{8} = 0, \\ 
    c_{1} - c_{9} = 0, \quad 2 c_{1} + c_{5} - c_{9} = 0, 
    \quad c_{5} + c_{7} - c_{9} = 0, \quad c_{2} - c_{10} = 0, \\
    c_{2} + 2 c_{6} - c_{10} = 0, \quad c_{2} - c_{10} = 0, 
    \quad c_{4} - c_{8} - c_{10} + c_{11} = 0, 
    \quad c_{2} - c_{8} + c_{11} = 0, \\
    c_{4} - c_{10} - c_{12} = 0, 
    \quad c_{11} - c_{12} = 0, \quad  c_{2} - c_{12} = 0, 
    \quad c_{11} - c_{12} = 0.
  \end{gathered}
\end{equation}
%
\vskip 1pt
\noindent
\step{Solve the linear system and build the recursion operator}
\vskip 3pt
\noindent
Solve the linear system and substitute the constants into the candidate 
recursion operator.  
For (\ref{dlw}), we find 
\begin{equation}
  c_1 = c_3 = c_5 = c_6 = c_7 = c_9 = 0, \quad 
  2 c_2 = c_4 = 2 c_{10} = 2 c_{11} = 2 c_{12} = c_8,
\end{equation}
so taking $c_8 = 1$ gives 
\begin{equation}
\label{recoperdlw}
  \R = 
  \begin{pmatrix}
    \frac{1}{2} v I \quad & u I + \frac{1}{2} u_x D_x^{-1} \\
    I \quad & \frac{1}{2} v I + \frac{1}{2} v_x D_x^{-1}
  \end{pmatrix}.
\end{equation}
In~\cite{Wang98} this recursion operator was obtained as the composition 
of the cosymplectic and symplectic operators of (\ref{dlw}).

Starting from $G^{(1)},$ repeated application of (\ref{recoperdlw}) 
generates an infinite number of generalized symmetries of (\ref{dlw}), 
establishing its completely integrable.  

%
%
\section{Other Software Packages}
\label{sec:otherSoftware}

There has been little work on using computer algebra methods to find 
and test recursion operators.
In 1987, Fuchssteiner et al.\ \cite{Fuchssteiner87} wrote \emph{PASCAL}, 
\emph{Maple}, and \emph{Macsyma} codes for testing recursion operators.  
While these packages could verify if a recursion operator is correct, 
they were unable to either generate the form of the operator or test 
a candidate recursion operator with undetermined coefficients.  
Bilge~\cite{Bilge93} did substantial work on finding recursion operators 
interactively with \emph{REDUCE}.  

Sanders and Wang~\cite{Wang98,Sanders98} wrote \emph{Maple} 
and \emph{Form} codes to aid in the computation of recursion operators.  
Their software was used to compute the symplectic, cosymplectic, as well 
as recursion operators of the 39 PDEs listed in \cite{Wang98} and
\cite{Wang02}.

Recently, Meshkov~\cite{Meshkov00} implemented a package in \emph{Maple} 
for investigating complete integrability from the geometric perspective.
If the zero curvature representation of the system is known, 
then his software package can compute the recursion operator.  
To our knowledge, our package 
\texttt{PDE\-Re\-cur\-sion\-Op\-er\-a\-tor.m}~\cite{recursioncode05} 
is the only fully automated software package for computing and testing 
polynomial recursion operators of polynomial evolution equations.  

%
%
\section{Additional Examples}
\label{sec:additionalExamples}

%
%
\subsection{The Nonlinear Schr\"odinger Equation}
\label{subsec:nls}

For convenience, we write the standard nonlinear Schr\"odinger equation (NLS),
\begin{equation}
\label{NLSeq}
  i u_t + u_{xx} + 2|u|^2u = 0,
\end{equation}
as the system of two real equations,  
\begin{equation}
  \label{nlsSys}
  \begin{gathered}
    u_t = - ( u_{xx} + 2 u^2 v ), \\
    v_t = v_{xx} + 2 u v^2,
  \end{gathered}
\end{equation}
where $v = \bar{u}$ and $i$ has been absorbed in $t.$

To determine the weights, we assume $W(u) = W(v)$ so that (\ref{nlsSys}) 
has dilation symmetry
$(t,x,u,v) \to (\lambda^{-2} t, \lambda^{-1} x, \lambda u, \lambda v).$  
Hence, $W(u) \!=\! W(v) \!=\! 1, W(D_t) \!=\! 2,$ and $W(D_x) \!=\! 1,$ 
as usual.
The first densities and symmetries are 
\begin{gather}
  \rho^{(1)} = uv, \qquad \rho^{(2)} = u_xv, \\
  \vec{G}^{(1)} = 
     \begin{pmatrix} 
            u \\ 
            -v 
     \end{pmatrix}, \qquad 
  \vec{G}^{(2)} = 
      \begin{pmatrix} 
            u_x \\ 
            v_x 
      \end{pmatrix}.
\end{gather}
Thus, $\rank\,\R_{ij} = 1, \,i,j = 1,2,$ and the candidate local operator is 
\begin{equation}
  \R_0 = 
  \begin{pmatrix}
    c_{1}D_x + ( c_{2}u + c_{3}v)I \quad & c_{4}D_x + ( c_{5}u + c_{6}v)I \\
    c_{7}D_x + ( c_{8}u + c_{9}v)I \quad & c_{10}D_x + ( c_{11}u + c_{12}v) I
  \end{pmatrix}.
\end{equation}
The candidate non-local operator is 
\begin{equation}
  \R_1 = \vec{G}^{(1)} D_x^{-1} \otimes \Euler{\vec{u}}(\rho^{(1)}) = 
    \begin{pmatrix} 
      -c_{13}uD_x^{-1}v I \quad & -c_{14}uD_x^{-1}u I \\
      c_{15}vD_x^{-1}v I \quad & c_{16}vD_x^{-1}u I
    \end{pmatrix}.
\end{equation}
Substituting $\R = \R_0 + \R_1$ into (\ref{defining}), solving for the 
undetermined coefficients, and setting $c_{16} = -2,$ we find 
\begin{equation}
  \label{nlsRec}
  \R = 
    \begin{pmatrix}
      D_x + 2 u D_x^{-1}v I \quad & 2 u D_x^{-1}u I \\
      -2 v D_x^{-1}v I \quad & -D_x - 2 v D_x^{-1}u I
    \end{pmatrix}.
\end{equation}
%
Starting from ``seed" $G^{(1)},$ the generalized symmetries can be constructed
sequentially using (\ref{sequentialsymmetries}).
This establishes the complete integrability of (\ref{nlsSys}).

In \cite{Wang98}, Wang split (\ref{NLSeq}) into an alternate system of two 
real equations by setting $u = v + i w.$ 
Using the cosymplectic and symplectic operators of that system, she obtained
a recursion operator which is equivalent to (\ref{nlsRec}).

%
%
\subsection{The Burgers' Equation}
\label{subsec:burgers}

Consider the Burgers' equation~\cite{Olver93},
\begin{equation}
  \label{burgers}
  u_t = u u_x + u_{xx},
\end{equation}
which has the dilation symmetry 
$(t,x,u) \to (\lambda^{-2} t, \lambda^{-1} x, \lambda u),$ or
$W(D_t) = 2,$ $ W(u) = 1,$ with $W(D_x) = 1.$
For (\ref{burgers}), 
\begin{equation}
  \rho^{(1)} = u, \qquad
  G^{(1)} = u_x, \qquad G^{(2)} = uu_x + u_{xx}.
\end{equation}
Assuming that $g = 1,$ the candidate recursion operator of $\rank\,\R = 1$ is 
\begin{equation}
  \R = c_1 D_x + c_2 u I + c_3 G^{(1)} D_x^{-1} \Euler{u}(\rho^{(1)})
    = c_1 D_x + c_2 u I + c_3 u_x D_x^{-1}.
\end{equation}
Using the defining equation (\ref{defining}), we determine that 
$c_1 = 2 c_3$ and $c_2 = c_3.$  
Taking $c_3 = \frac{1}{2},$ gives the recursion operator reported 
in~\cite{Olver77},
\begin{equation}
\label{Rstandardburgers}
  \R = D_x + \frac{1}{2} \left( u I + u_x D_x^{-1} \right) 
     = D_x + \frac{1}{2} D_x \left( u D_x^{-1} \right).
\end{equation}  
As expected, starting from $G^{(1)},$ one computes $G^{(2)} = \R G^{(1)},$ 
$G^{(3)} = \R G^{(2)} = \R^2 G^{(1)},$ etc., confirming that (\ref{burgers}) 
is completely integrable. 

The Burgers' equation also has the recursion operator~\cite{Olver93},
\begin{equation}
  \label{Rtburgers}
  \tilde{\R} = t \R + \frac{1}{2} \left( x I + D_x^{-1} \right)
    = t D_x + \frac{1}{2} \left( t u + x \right) I 
      + \frac{1}{2} \left( t u_x + 1 \right) D_x^{-1},
\end{equation}
which explicitly depends on $x$ and $t.$  
Using $W(t) = -2, W(x) = W(D_x^{-1}) = -1 $ and $W(u) = 2,$ 
one can readily verify that each term in (\ref{Rtburgers}) has rank $-1.$

To find recursion operators like (\ref{Rtburgers}), which depend explicitly 
on $x$ and $t,$ we can again use scaling symmetries to build $\tilde{\R}.$  
However, one must select the maximum degree for $x$ and $t.$ 
For instance, for degree 1 the coefficients in the recursion operator will
at most depend on $x$ and $t$ (but not on $x t$, $x^2,$ or $t^2$ which are 
quadratic).  
To control the highest exponent in $x$ and $t,$ in the code the user can set 
\verb|MaxExplicitDependency| to any non-negative integer value 
(see Appendix~\ref{sec:softwareUsage}).

With \verb|MaxExplicitDependency -> 1|, the candidate local operator then is 
\begin{equation}
  \tilde{\R}_0 = c_1 t D_x + \left( c_2 x + c_3 t u \right) I.
\end{equation}
The first symmetries that explicitly depend on $x$ and $t$ (of degree 1) are
\begin{equation}
  \tilde{G}^{(1)} = 1 + t u_x, \qquad 
  \tilde{G}^{(2)} = \frac{1}{2} \left( u + x u_x \right) + tuu_x + tu_{xx}.
\end{equation}
Thus, the candidate non-local operator is 
\begin{equation}
  \tilde{\R}_1 = c_4 \tilde{G}^{(1)} D_x^{-1} \Euler{u}(\rho^{(1)})
    = c_4 \left( tu_x + 1 \right) D_x^{-1}.
\end{equation}
Requiring that $\tilde{R} = \tilde{\R}_0 + \tilde{\R}_1$ satisfies the 
defining equation (\ref{defining}), 
next solving for the constants $c_1$ through $c_4,$ and finally setting 
$c_4 = \frac{1}{2},$ yields the recursion operator (\ref{Rtburgers}).  
%
%
Using (\ref{Rtburgers}), one can construct an additional infinite sequence 
of generalized symmetries. 
Furthermore, $\tilde{G}^{(2)} = \tilde{\R} G^{(1)},$
$\tilde{G}^{(3)} = \tilde{\R} G^{(2)} = \tilde{\R} \R G^{(1)},$ etc.
Connections between $\R$ and ${\tilde{\R}}$ and their symmetries are 
discussed in \cite{Olver93} and \cite{Wang98}.

%
%
\subsection{The Drinfel'd-Sokolov-Wilson Equation}
\label{subsec:drinfeld}

Consider the Drinfel'd-Sokolov-Wilson system~\cite{Ablowitz91,Hirota86}, 
\vspace{-1mm}
\noindent
\begin{equation}
  \label{dsw}
  \begin{aligned}
    u_t & = 3 v v_x, \\
    v_t & = 2 u v_x + u_x v + 2 v_{3x},
  \end{aligned}
\end{equation}
\vskip 0.1pt
\noindent
which has static soliton solutions that interact with moving solitons 
without deformation.  
The scaling symmetry for (\ref{dsw}) is 
$(t,x,u,v) \to (\lambda^{-3} t, \lambda^{-1}x, \lambda^2 u, \lambda^2 v).$  
Expressed in weights, $W(D_t) = 3, W(D_x) = 1,$ and $W(u) = W(v) = 2.$ 
%
The first few conserved densities and generalized symmetries are 
\vspace{-2mm}
\noindent
\begin{gather}
  \rho^{(1)} = u, \qquad \rho^{(2)} = v^2, \qquad 
  \rho^{(3)} = \frac{4}{27} u^3 - \frac{2}{3} uv^2 -\frac{1}{9} u_x^2 + v_x^2,
  \\
\label{G1G2Drinfeld}
  \vec{G}^{(1)} = 
      \begin{pmatrix} 
            u_x \\ 
            v_x 
      \end{pmatrix}, \qquad 
  \vec{G}^{(2)} = 
    \begin{pmatrix} 
           3 vv_x \\ 
           u_x v + 2 uv_x + 2 v_{3x} 
    \end{pmatrix}, 
\end{gather}
\vspace{-2mm}
\noindent
and 
\vspace{-2mm}
\noindent
\begin{equation}
\label{G3Drinfeld}
  \vec{G}^{(3)} \!\!=\!\!
    \begin{pmatrix} 
    -\! 10 u^2 u_x \!+\! 15 v^2 u_x \!+\! 30 u v v_x -\! 25 u_x u_{2x} 
    \!+\! 45 v_x v_{2x} -\! 10 u u_{3x} \!+\! 30 v v_{3x} -\! 2 u_{5x} \\ 
    10 u^2 v_x \!+\! 15 v^2 v_x \!+\! 10 u v u_x \!+\! 45 u_x v_{2x} 
    \!+\! 35 v_x u_{2x} \!+\! 30 u v_{3x} \!+\! 10 v u_{3x} \!+\! 18 v_{5x}
    \end{pmatrix}\!.
\end{equation}
\vspace{-2mm}
\noindent
Despite the fact that 
\vspace{-2mm}
\noindent
\begin{equation}
\label{ranksymmetriesdrinfeldetal}
  \begin{gathered}
  \rank\,\vec{G}^{(1)} = 
  \begin{pmatrix} 
  3 \\ 3 
  \end{pmatrix}, \qquad
  \rank\,\vec{G}^{(2)} = 
  \begin{pmatrix} 
  5 \\ 5 
  \end{pmatrix}, 
  \end{gathered}
\end{equation}
\vskip 0.1pt
\noindent
we can not take $g \!=\! 1$ or $2.$
Surprisingly, for (\ref{dsw}) we must set $g \!=\! 3$ and 
$\rank\,\R_{ij} \!=\! 6,\, i,j = 1,2.$  
So, the candidate local operator has elements involving $D_x^6.$  
For example, 
\vspace{-2mm}
\noindent
\begin{multline}
  (\R_0)_{11} = c_1 D_x^{6} + \left( c_2 u + c_5 v \right) D_x^{4} 
  + \left( c_8 u_{x} + c_{10}v_{x} \right) D_x^{3} 
  + \left( c_3 u^{2} + c_6 v^{2} + c_{12}u_{2 x} \right. \\
  \left. + c_{13}v_{2 x} + c_{18}uv \right) D_x^{2} 
  + \left( c_{14}u_{3 x} + c_{15}v_{3 x} + c_{20}uu_{x} + c_{21}uv_{x} 
  + c_{25}vu_{x} + c_{26}vv_{x} \right) D_x \\
  + \left( c_4 u^{3} + c_7 v^{3} + c_{27}vu_{2 x} 
  + c_{28}vv_{2 x} + c_{29}u_{x}v_{x} 
  + c_9 u_{x}^{2} + c_{11}v_{x}^{2} + c_{16}u_{4 x} \right. \\ 
  \left. + c_{17}v_{4 x} + c_{19}uv^{2} + c_{22}uu_{2 x} + c_{23}uv_{2 x} 
  + c_{24}u^{2}v \right) I.
\end{multline}
\vspace{-2mm}
\noindent
The candidate non-local operator is 
\begin{equation}
  \R_1 = 
   \sum_{i=1}^4 \vec{G}^{(i)} D_x^{-1} \otimes \Euler{\vec{u}}( \rho^{(5-i)} )
   = \begin{pmatrix} 
     (\R_1)_{11} \quad & (\R_1)_{12} \\ (\R_1)_{21} \quad & (\R_1)_{22} 
     \end{pmatrix},
\end{equation}
where 
\begin{align}
(\R_1)_{11} & = \notag 
- \left( \frac{1}{9}c_{117}u_{5 x}
   + \frac{25}{18}c_{118}u_{x}u_{2 x}
   + \frac{5}{9}c_{119}uu_{3 x}
   + \frac{5}{9}c_{120}u^{2}u_{x} \right. \\ \notag &
\left. - \frac{5}{6}c_{121}v^{2}u_{x}
       - \frac{5}{3}c_{122}vv_{3 x}
       - \frac{5}{2}c_{123}v_{x}v_{2 x} \right) D_x^{-1} 
- \frac{2}{3}c_{124}u_{x}D_x^{-1}v^{2} \\ &
+ \frac{2}{9}c_{125}u_{x}D_x^{-1}u_{2x} 
+ \frac{4}{9}c_{126}u_{x}D_x^{-1}u^{2} 
+ \frac{5}{3}c_{127}uvv_{x}D_x^{-1}, \\
(\R_1)_{12} & = 
- 2c_{128}u_{x}D_x^{-1}v_{2 x} 
- \frac{4}{3}c_{129}u_{x}D_x^{-1}uv
+ 3c_{130}vv_{x}D_x^{-1}v, \\
(\R_1)_{21} & = \notag 
\left( c_{131}v_{5 x}
     + \frac{5}{9}c_{132}u^{2}v_{x}
     + \frac{5}{9}c_{133}vu_{3 x}
     + \frac{5}{6}c_{134}v^{2}v_{x} \right. \\ \notag &
     \left. + \frac{5}{3}c_{135}uv_{3 x}
     + \frac{35}{18}c_{136}v_{x}u_{2 x}
     + \frac{5}{2}c_{137}u_{x}v_{2 x} \right) D_x^{-1} 
- \frac{2}{3}c_{138}v_{x}D_x^{-1}v^{2} \\ &
+ \frac{2}{9}c_{139}v_{x}D_x^{-1}u_{2 x} 
+ \frac{4}{9}c_{140}v_{x}D_x^{-1}u^{2} 
+ \frac{5}{9}c_{141}uvu_{x}D_x^{-1}, \\
(\R_1)_{22} & = \notag 
- 2c_{142}v_{x}D_x^{-1}v_{2 x} 
+ 2c_{143}v_{3 x}D_x^{-1}v 
+ 2c_{144}uv_{x}D_x^{-1}v \\ &
- \frac{4}{3}c_{145}v_{x}D_x^{-1}uv
+ c_{146}vu_{x}D_x^{-1}v. 
\end{align}
The terms in (\ref{defining}) fill 160 pages and grouping like terms 
results in a system of 508 linear equations for $c_i.$  
Solving these linear equations and setting $c_{146} = -9$ gives the 
recursion operator 
\begin{equation}
  \R = \begin{pmatrix} 
    (\R)_{11} \quad & (\R)_{12} \\ (\R)_{21} \quad & (\R)_{22}
  \end{pmatrix}
\end{equation}
with 
\vspace{-1mm}
\noindent
\begin{multline}
  (\R)_{11} = D_x^{6} + 6 u D_x^{4} + 18 u_{x} D_x^{3} 
  + \Big( 9u^{2} - 21v^{2} + \frac{49}{2}u_{2 x} \Big) D_x^{2}  
  + \Big( 30uu_{x} - 75vv_{x} \\ + \frac{35}{2}u_{3 x}\Big) D_x 
  + \Big( 4u^{3} - 12uv^{2} + \frac{41}{2}uu_{2 x} + \frac{13}{2}u_{4 x} 
  + \frac{69}{4}u_{x}^{2} - \frac{111}{2}vv_{2 x} \\ 
  - \frac{141}{4}v_{x}^{2} \Big) I 
+ \Big( 5u^{2}u_{x} + 5uu_{3 x} - 15uvv_{x} 
- 15vv_{3 x} - \frac{15}{2}v^{2}u_{x} + \frac{25}{2}u_{x}u_{2 x} \\
- \frac{45}{2}v_{x}v_{2 x} + u_{5 x} \Big) D_x^{-1} 
+ \frac{1}{2}u_{x} D_x^{-1} u_{2 x} I - \frac{3}{2} u_{x} D_x^{-1} v^{2} I 
+ u_{x} D_x^{-1} u^{2} I,
\end{multline}
\vspace{-2mm}
\noindent
\begin{multline}
  (\R)_{12} = - 42v D_x^{4} - 51v_{x} D_x^{3} 
- \Big( 48uv + \frac{63}{2}v_{2 x} \Big) D_x^{2} 
- \Big( 33uv_{x} + 60vu_{x} + \frac{21}{2}v_{3 x}\Big) D_x \\ 
- \Big( 18v^{3} + 15u_{x}v_{x} + 6u^{2}v 
+ \frac{15}{2}uv_{2 x} + \frac{39}{2}vu_{2 x} + \frac{3}{2}v_{4 x} \Big) I \\
- 27vv_{x} D_x^{-1} v I - 3 u_{x} D_x^{-1} uv I 
- \frac{9}{2} u_{x} D_x^{-1}v_{2 x} I, 
\end{multline}
\vspace{-2mm}
\noindent
\begin{multline}
  (\R)_{21} = - 14v D_x^{4} - 67v_{x} D_x^{3} 
- \Big( 16uv + \frac{243}{2} v_{2 x}\Big) D_x^{2} 
- \Big( 18vu_{x} + 53uv_{x} \\ 
+ \frac{219}{2}v_{3 x} \Big) D_x 
- \Big( 46u_{x}v_{x} + 2u^{2}v + 6v^{3} 
+ \frac{99}{2}uv_{2 x} + \frac{99}{2}v_{4 x} + \frac{27}{2}vu_{2 x} \Big) I \\
- \Big( 15uv_{3 x} + 5u^{2}v_{x} + 5uvu_{x} + 5vu_{3 x} + 9v_{5 x} 
+ \frac{15}{2}v^{2}v_{x} + \frac{35}{2}v_{x}u_{2 x} \\ 
+ \frac{45}{2}u_{x}v_{2 x} \Big) D_x^{-1} 
+ \frac{1}{2}v_{x} D_x^{-1} u_{2 x} I
- \frac{3}{2} v_{x} D_x^{-1} v^{2} I + v_{x} D_x^{-1} u^{2} I,
\end{multline}
\vskip 0.1pt
and 
\vspace{-3mm}
\noindent
\begin{multline}
  (\R)_{22} = - 27 D_x^{6} - 54 u D_x^{4} - 108 u_{x} D_x^{3} 
- \Big( 27u^{2} + 33v^{2} + \frac{243}{2}u_{2 x} \Big) D_x^{2} \\ 
- \Big( 54uu_{x} + 105vv_{x} + \frac{135}{2}u_{3 x} \Big) D_x 
- \Big( 24uv^{2} + \frac{27}{2}uu_{2 x} + \frac{27}{4}u_{x}^{2} \\ 
+ \frac{147}{2}vv_{2 x} + \frac{27}{2}u_{4 x} + \frac{201}{4}v_{x}^{2} \Big) I 
- 9 \Big( 2 uv_{x} + 2 v_{3 x} + vu_{x} \Big) D_x^{-1} v I \\ 
- 3v_{x} D_x^{-1} uv I - \frac{9}{2} v_{x} D_x^{-1} v_{2 x} I
\end{multline}
\vskip 0.1pt
\noindent
%
%
%
This recursion operator can also be computed \cite{Wang98} by composing the 
cosymplectic and symplectic operators of (\ref{dsw}).
Since $g \!=\! 3$ the symmetries are not generated via 
(\ref{sequentialsymmetries}).
Instead, there are three seeds, $G^{(1)}, G^{(2)},$ and $G^{(3)}$ given
in (\ref{G1G2Drinfeld}) and (\ref{G3Drinfeld}).
Using (\ref{nonsequentialsymmetries}), from $G^{(1)}$ one obtains 
$G^{(4)} = \R G^{(1)},$ 
$G^{(7)} = \R G^{(4)} = \R^2 G^{(1)},$ and so on.
From $G^{(2)},$ upon repeated application of $\R,$ one gets 
$G^{(5)}, G^{(8)},$ etc., whereas $G^{(3)}$ generates $G^{(6)}, G^{(9)},$ etc.
Thus, the recursion operator generates a threefold infinity of generalized 
symmetries, confirming that (\ref{dsw}) is completely integrable. 

This example illustrates the importance of computer algebra software 
in the study of integrability, in particular, for the computation of 
recursion operators. 
The length of the computations makes it impossible to compute the 
recursion operators for all but the simplest systems by hand.  

%
%
\vspace{-3mm}
\noindent
\section{Conclusions and Future Work}
\label{sec:conclusions}

To our knowledge, no one has ever attempted to fully automate an algorithm 
for finding or testing recursion operators.  
The commutative nature of computer algebra systems makes it a non-trivial 
task to efficiently implement the non-commutative rules needed for working 
with integro-differential operators. 

Based on our recursion operator algorithm and it implementation in 
\texttt{PDE\-Re\-cur\-sion\-Op\-er\-a\-tor.m}, a large class of nonlinear PDEs 
can be tested for complete integrability in a straightforward manner.  
Currently our code computes polynomial recursion operators for polynomial 
PDEs (with constant coefficients) which can be written in evolution form,
$\vec{u}_t = \vec{F}(\vec{u},\vec{u}_x, \dotsc, \vec{u}_{mx}).$ 

With the tools developed for finding and testing recursion operators, 
it would be possible to extend the algorithm to find master symmetries 
as well as cosymplectic, symplectic and conjugate recursion operators.
A symplectic operator maps (generalized) symmetries into cosymmetries, 
while a cosymplectic operator maps cosymmetries into generalized symmetries.  
Hence, the recursion operator for a system is the composition of the 
cosymplectic operator and the symplectic operator of the system.  
A conjugate recursion operator maps conserved densities of lower order to 
conserved densities of higher order.  
The master symmetry can be used to generate an infinite hierarchy of 
time-dependent generalized symmetries.  
%
It would be worthwhile to add to our {\it Mathematica} code an automated 
test of the ``hereditary" condition \cite{Fuchssteiner87} for 
recursion operators.  

%
%
\vspace{-3mm}
\noindent
\section*{Acknowledgements}
\label{acknowledgements}
This material is based upon work supported by the National Science Foundation 
(NSF) under Grant Nos.\ CCF-0830783 and CCR-9901929.  
This work was also partially supported by a NDSEG Fellowship awarded to DB.
Any opinions, findings, and conclusions or recommendations expressed in this
material are those of the authors and do not necessarily reflect the views 
of NSF. 

WH thanks Jan Sanders (Free University Amsterdam, The Netherlands) and 
Jing Ping Wang (University of Kent, Canterbury, UK) for many valuable 
discussions.

%
%

\vspace{-9mm}
\noindent
\appendix

%
%
\section{Using the Software Package \texttt{PDERecursionOperator.m}}
\label{sec:softwareUsage}

The package \texttt{PDE\-Re\-cur\-sion\-Op\-er\-a\-tor.m} has been tested 
with Mathematica 4.0 through 7.0 using more than 30 PDEs.  
The Backus-Naur form of the main function (\verb|RecursionOperator|) is
\begin{tabbing}
$\qquad$ \= 
$\langle Main\, Function \rangle \quad \to \quad$ \=
$\texttt{RecursionOperator}[\langle Equations \rangle, 
\langle Functions \rangle,$ \+ \+ \\
$\qquad \langle Variables \rangle, \langle Parameters \rangle,
\langle Options \rangle]$ \- \\ 
$\langle Options \rangle \quad \to \quad $ \= 
$\verb|Verbose|\rightarrow \langle Bool \rangle \; | \; 
 \verb|WeightsVerbose|\rightarrow \langle Bool \rangle \; | $ \+ \\
$\verb|Gap| \rightarrow \langle Positive~Integer \rangle \; |$ \\
$\verb|MaxExplicitDependency| \rightarrow 
  \langle Nonnegative~Integer \rangle \; |$ \\
$\verb|RankShift| \rightarrow 
  \langle Integer \rangle \; |$ \\
$\verb|WeightRules| \rightarrow
  \langle List~of~Rules \rangle \; |$ \\
$\verb|WeightedParameters| \rightarrow
  \langle List~of~Weighted~Parameters\rangle \; |$ \\
$\verb|UnknownCoefficients| \rightarrow
  \langle Symbol \rangle $ \- \\
$\langle Bool \rangle \quad \to \quad \mathtt{True} \; | \; \mathtt{False}$ \\
$\langle List~of~Rules \rangle \quad \to \quad \mathtt{
\{ weight[u] \rightarrow} \langle Integer \rangle\mathtt{, 
  weight[v] \rightarrow}\langle Integer \rangle \mathtt{,... \}} $ \\
\end{tabbing}
\vskip 1pt
\noindent
When using a PC, place the packages 
\texttt{PDE\-Re\-cur\-sion\-Op\-er\-a\-tor.m} 
and \texttt{In\-var\-i\-ants\-Sym\-me\-tries.m} in a directory, 
for example, \verb|myDirectory| on drive \verb|C|.  
Start a \textit{Mathematica} notebook session and execute the commands:
\begin{verbatim}
In[1] := SetDirectory["C:\\myDirectory"];   (* Specify directory *)

In[2] := Get["PDERecursionOperator.m"]    (* Read in the package *)

In[3] := RecursionOperator[                 (* Burgers' equation *)
           D[u[x,t],t]==u[x,t]*D[u[x,t],x]+D[u[x,t],{x,2}], 
           u[x,t], {x,t}]
Out[3] = 
\end{verbatim}
\[ 
  \{\{\{2 C_3D_x + C_3uI + C_3u_{x}D_x^{-1}\}\}\} 
\]
We can find a recursion operator for Burgers' equation which explicitly 
depends on $x$ and $t$ (linearly) by using the option 
\verb|MaxExplicitDependency|:
\begin{verbatim}
In[4] := RecursionOperator[                 (* Burgers' equation *)
           D[u[x,t],t]==u[x,t]*D[u[x,t],x]+D[u[x,t],{x,2}], 
           u[x,t], {x,t}, MaxExplicitDependency -> 1]
Out[4] = 
\end{verbatim}
\[ 
  \{\{\{2 C_5tD_x + C_5(x + tu)I + C_5(1 + tu_{x})D_x^{-1} \}\}\} 
\]
\begin{verbatim}
In[5] := RecursionOperator[           (* Potential mKdV equation *)
           D[u[x,t],t]==1/3*D[u[x,t],x]^3+D[u[x,t],{x,3}],
           u[x,t], {x,t}, 
           WeightRules -> {weight[u] -> 1}, Gap -> 2]
\end{verbatim}
\[ 
  \{\{\{ 3 C_{19}D_x^{2} + (C_1 + 2 C_{19}u_x^2)I 
  - 2C_{19}u_{x}D_x^{-1}u_{2 x}I \}\}\} 
\]
In this example, 
we must use the \verb|WeightRules| option to fix the weights and the 
\verb|Gap| option to set $g = 2.$
\begin{verbatim}
In[6] := RecursionOperator[                  (* Diffusion system *)
           { D[u[x,t],t]==D[u[x,t],{x,2}]+v[x,t]^2, 
             D[v[x,t],t]==D[v[x,t],{x, 2}] }, 
           { u[x,t],v[x,t] }, {x,t}, 
           WeightRules -> {weight[u] -> weight[v]}, 
           RankShift -> -1 ]
\end{verbatim}
\[ 
\{\{\{C_5D_x, C_2D_x + C_5vD_x^{-1}\}, \{0, C_5D_x\}\}\} 
\]
In this system of equations, we again use the option \verb|WeightRules| 
to fix the weights.  
We also use the option \verb|RankShift| to force  \verb|RecursionOperator| 
to search for recursion operator of a lower than expected rank.

The option \verb|Verbose| prints out a trace of the calculations, 
while the option \verb|WeightsVerbose| prints out a trace of the calculation 
of the scaling symmetry. 
If one or more parameters have a weight, these weights can be specified using 
\verb|WeightedParameters|. 
The undetermined constants can be set to any variable using the option 
\verb|UnknownCoefficients| 
(the default is $C_i$).

%
%

\end{document}